\documentclass{article}

\usepackage{epsfig,graphics}
\usepackage{enumerate, xspace}
\usepackage{amssymb, amstext,amsmath}
\usepackage{color}
\usepackage{changebar,framed}
\usepackage{xspace}
\usepackage{algorithmic}

\newtheorem{definition}{Definition}[section]
\newtheorem{theorem}{Theorem}[section]
\newtheorem{example}{Example}[section]
\newtheorem{lemma}{Lemma}[section]
\newtheorem{remark}{Remark}[section]

\newtheorem{proposition}{Proposition}[section]
\newtheorem{corollary}{Corollary}[section]
\newtheorem{observation}{Observation}[section]
\newtheorem{conjecture}{Conjecture}[section]

\newcommand{\ab}{\allowbreak }
\newcommand{\splits}{\hfill\cr{}\hfill}
\newcommand{\predi}[1]{\text{\bf #1}\xspace}
\newcommand{\fol}{first-order\xspace }

\newcommand{\fo}{${\rm FO}$\xspace}

\newcommand{\foms}[1]{{\rm FO}({#1})\xspace}
\newcommand{\fom}[2]{{\rm FO}({#1}, {#2})\xspace}

\newcommand{\fobsigma}{\fom{$\{\betws\}$}{$\sigmapt$}\xspace}

\newcommand{\fobgenSTsigma}{\fom{$\{\betws,\ab \befores, \ab qcrsts \}$}{$\sigmaptST$}\xspace}

\newcommand{\fob}{\foms{$\{\betws\}$}} 

\newcommand{\fobST}{\foms{$\{\betwsCotemp,\ab \befores, \ab \eqcrsts \}$}}
\newcommand{\fobSTsigma}{\fom{$\{\betwsCotemp,\ab \befores, \ab \eqcrsts \}$}{$\sigmaptST$}\xspace}
\newcommand{\fod}{FO($\{\pos\}$)\xspace}

\newcommand{\betws}{\predi{Between}\xspace}
\newcommand{\betw}[3]{\predi{Between}(#1, \ab #2, \ab #3)\xspace}
\newcommand{\betwsCotemp}{\predi{Between}^\textrm{Cotemp}\xspace}
\newcommand{\betwCotemp}[3]{\predi{Between}^\textrm{Cotemp}(#1, \ab #2, \ab #3)\xspace}
\newcommand{\befores}{\predi{Before}}
\newcommand{\before}[2]{\predi{Before}(#1,\ab #2)\xspace}
\newcommand{\eqcrsts}{\predi{EqCr}^{ST}\xspace}
\newcommand{\eqcrst}[6]{\predi{EqCr}^{ST}(#1, \ab #2, \ab #3, \ab #4, \ab #5, \ab #6)\xspace}
\newcommand{\pos}{\predi{PartOf}}
\newcommand{\po}[2]{\predi{PartOf}(#1, #2)}
\newcommand{\pt}[1]{\predi{Point}(#1)}

\newcommand{\st}{{\rm spatio-temporal}\xspace}

\newcommand{\ie}{\emph{i.e.}}
\newcommand{\sa}{semi-algebraic\xspace }

\newcommand{\cotemp}{co-temporal\xspace}
\newcommand {\fok}{FO\xspace }

\newcommand{\ndim}[1]{${#1}$-{\rm dimensional}\xspace}
\newcommand{\R}{\mathbb{R}}

\newcommand{\Rn}[1]{\mathbb{R}^{#1}}
\newcommand{\RnR}[1]{(\mathbb{R}^{#1}\times\mathbb{R})}
\newcommand{\Rnm}[2]{{(\mathbb{R}^{#1})}^{#2}}
\newcommand{\RnRm}[2]{{(\mathbb{R}^{#1}\times\mathbb{R})}^{#2}}
\newcommand{\Rnmk}[3]{{({(\mathbb{R}^{#1})}^{#2})}^{#3}}
\newcommand{\RnRmk}[3]{{({(\mathbb{R}^{#1}\times\mathbb{R})}^{#2})}^{#3}}

\newcommand{\stb}[1]{{\cal S\!\!\!\!T}\!^{#1}}

\newcommand{\tr}[1]{\vartriangle_{#1}}
\newcommand{\trc}[1]{T_{#1}}
\newcommand{\trST}[1]{\vartriangle_{#1}^{st}}
\newcommand{\trcST}[1]{T_{#1}^{st}}

\newcommand{\vect}[1]{\mathbf{#1}}

\newcommand{\structTr}{{\cal D}}
\newcommand{\structTrST}{{\cal D}^{st}}
\newcommand{\structPt}{{\cal S}}
\newcommand{\structST}{{\cal S\!\!\!\!T}}

\newcommand{\sigmaST}{\sigma^{st}}
\newcommand{\RST}{R^{st}}
\newcommand{\sigmatr}{\hat{\sigma}}
\newcommand{\Rtr}{\hat{R}}
\newcommand{\sigmatrST}{{\hat{\sigma}}^{st}}
\newcommand{\RtrST}{{\hat{R}}^{st}}
\newcommand{\sigmapt}{\dot{\sigma}}
\newcommand{\Rpt}{\dot{R}}
\newcommand{\sigmaptST}{{\dot{\sigma}}^{st}}
\newcommand{\RptST}{{\dot{R}}^{st}}

\newcommand{\sigPolyNoBr}{$+$, $\times$, $<$, $0$, $1$\xspace }

\newcommand{\betwtr}[3]{\predi{Between}_{\Delta}(#1, #2, #3)}

\newcommand{\posCotemp}{\predi{PartOf}^\textrm{Cotemp}}
\newcommand{\poCotemp}[2]{\predi{PartOf}^\textrm{Cotemp}(#1, \ab #2)}

\newcommand{\beforetrs}{\predi{Before}_{\vartriangle}}
\newcommand{\beforetr}[2]{\predi{Before}_{\vartriangle}(#1,#2)}

\newcommand{\posnk}[2]{\predi{PartOf}^{({#1},{#2})}}

\newcommand{\qtr}{\hat{Q}}
\newcommand{\qpt}{\dot{Q}}

\def\squareforqed{\hbox{\rlap{$\sqcap$}$\sqcup$}}
\def\qed{\ifmmode\squareforqed\else{\unskip\nobreak\hfil
\penalty50\hskip1em\null\nobreak\hfil\squareforqed
\parfillskip=0pt\finalhyphendemerits=0\endgraf}\fi}

\begin{document}

\title{A triangle-based logic for affine-invariant querying of spatial and spatio-temporal data}

\author{\emph{Sofie Haesevoets}, \emph{Bart Kuijpers}\footnote{Corresponding author: Hasselt University,
Theoretical Computer Science, B-3590 Diepenbeek, Belgium, {\tt
bart.kuijpers@uhasselt.be}}\\ Hasselt University}

 \maketitle

\begin{abstract}
In spatial databases, incompatibilities often arise due to different choices of origin or unit of
measurement (e.g., centimeters versus inches). By representing and querying the data in an
affine-invariant manner, we can avoid these incompatibilities.

In practice, spatial (resp., spatio-temporal) data is often represented as a finite union of
triangles (resp., moving triangles). As two arbitrary triangles are equal up to a unique affinity
of the plane, they seem perfect candidates as basic units for an affine-invariant query language.

We propose a so-called ``triangle logic'', a query language that is affine-generic and has
triangles as basic elements. We show that this language has the same expressive power as the
affine-generic fragment of first-order logic over the reals on triangle databases. We illustrate
that the proposed language is simple and intuitive. It can also serve as a first step towards a
``moving-triangle logic'' for spatio-temporal data.
\end{abstract}

\section{Introduction and summary of the results}\label{sec-introduction}

In the area of spatial database research, a lot of attention has been focused on affine invariance
of both data and queries. The main purpose of studying affine invariance is to obtain methods and
techniques that are not affected by affine transformations of the input spatial data. This means
that a particular choice of origin or some artificial choice of unit of measure (e.g., inches
versus centimeters) has no effect on the final result of the method or query.

In computer vision, the so-called {\em weak perspective assumption}~\cite{weakpersp} is widely
adopted. When an object is repeatedly photographed under different camera angles, all the different
images are assumed to be affine transformations of each other. This assumption led to the need for
affine-invariant similarity measures between pairs of
pictures~\cite{sym-diff,pattern-hausdorff,geom-hashing}. In computer graphics, affine-invariant
norms and triangulations have been studied~\cite{nielson}. In the field of spatial and
spatio-temporal constraint databases~\cite{cdbook,reveszbook}, affine-invariant query
languages~\cite{ghk-01,gvv-jcss} have been proposed. In
Section~\ref{sec-relatedwork-preliminaries}, we will go into more detail about the affine-invariant
language for spatial constraint data proposed by Gyssens, Van den Bussche and Van
Gucht~\cite{gvv-jcss}. Affinities are one of the transformation groups proposed at the introduction
of the concept of ``genericity of query languages'' applied to constraint
databases~\cite{pvv-pods}. Also various subgroups of the affinities~\cite{pvv-pods} such as
isometries, similarities, \ldots\ and supergroups of the affinities such as topology preserving
transformations~\cite{pkv_icdt97,psv_topo} have been studied in the same context.

If we now focus on the representation of two-dimensional spatial data, we see that, in practice,
two-dimensional figures are approximated often as a finite union of triangles. In geographic
information systems, Triangulated Irregular Networks (TIN)~\cite{cdbondkdlk} are often used. In
computer graphics, data is approximated by triangular meshes (e.g.,~\cite{comp-geom}). Also, for
spatio-temporal databases, ``parametric moving triangle''-based models have been proposed and
studied~\cite{cz-00,amai03}.

Remark that two arbitrary triangles are indistinguishable up to an affinity of the plane. Indeed,
each triangle in the plane can be mapped to each other triangle in the plane by a unique
affinity.\footnote{This is only true if the triangle is not degenerated, i.e., no corner points
coincide. Otherwise, there are more such affinities.}

The combination of the need for affine-invariance, the representation of data by means of triangles
in practice, and the fact that triangles itself are an affine-invariant concept, led to the idea of
introducing a query language based on triangles. If the data is represented as a collection of
triangles, why should one reason about it as a collection of points~\cite{gvv-jcss}, or, even
indirectly, by means of coordinates (as is the case for the classical spatial constraint language,
first-order logic over the reals)? We consider first-order languages, in which variables are
interpreted to range over triangles, both spatial and spatio-temporal.

We propose a new, first-order query language that has triangles as basic elements. We show that
this language has the same expressive power as the affine-invariant segment of the queries in
first-order logic over the reals on triangle databases. Afterwards, we give some examples
illustrating the expressiveness of our language. We also address the notion of safety of triangle
queries. We show that it is undecidable whether a specific triangle query returns a finite output
on finite input. It is, however, decidable whether the output of a query on a particular finite
input database can be represented as a finite union of triangles. We show that we can express this
finite representation in our triangle language. Afterwards, we extend our results to the case of
\st triangles, \ie, triples of \cotemp points in $\RnR{2}$.

\section{Related work and preliminaries}\label{sec-relatedwork-preliminaries} The idea that the result of a query on some spatial input database should be invariant
under some group of spatial transformations, was first introduced by Paredaens, Van den Bussche and
Van Gucht~\cite{pvv-pods}. In a follow-up article, Gyssens, Van den Bussche and Van
Gucht~\cite{gvv-jcss} proposed several first-order query languages, invariant under group of the
affinities or some subgroup thereof. In these languages, variables are assumed to range over points
in some real space $\R^n$ ($\R$ is the set of real numbers), rather than over real numbers
(coordinates of such points). For the group of the affinities, the point language with only one
predicate that expresses {\em betweenness} of points, was shown to have the same expressivity as
the affine-invariant fragment of first-order logic over the reals, on point databases. We will use
this result to prove the expressiveness of our triangle-based logic. Therefore, we will recall some
definitions from the article from Gyssens, Van den Bussche and Van Gucht~\cite{gvv-jcss}. All
definitions listed in this section can be found there.

We start with the well-known definition of a constraint database, or semi-algebraic database, as
this is the general setting which we will be working in.  

\begin{definition}\rm
A \emph{semi-algebraic relation in $\R^n$\/} is a subset of $\R^n$ that can be described as a
Boolean combination of sets of the form
$$\{ (x_1, x_2, \ldots, x_n) \in \R^n \mid p(x_1, x_2, \ldots, x_n) > 0 \},$$ with $p$ a polynomial with integer coefficients in the real variables
$x_1, x_2, \ldots, x_n$. \qed
\end{definition}

In mathematical terms, semi-algebraic relations are known as \emph{semi-algebraic
sets}~\cite{bcr-real}.

We also call a semi-algebraic relation in $\R^n$ \emph{a semi-algebraic relation of arity $n$}. A
semi-algebraic database is essentially a finite collection of semi-algebraic relations. We give the
definition next.

\begin{definition}\label{sad-def}\rm
\medskip\par
A \emph{(semi-algebraic) database schema $\sigma$} is a finite set of relation names, where each relation name $R$ has an arity associated to it,
which is a natural number and which is denoted by $ar(R)$.

Let $\sigma$ be a database schema.  A \emph{semi-algebraic database over $\sigma$} is a structure
$\cal{D}$ over $\sigma$ with domain $\R$ such that, for each relation name $R$ of $\sigma$, the
associated relation  $R^{\cal{D}}$ in ${\cal D}$ is a semi-algebraic relation of arity $ar(R)$.
\qed
\end{definition}

\begin{example}\rm\label{sadb}
Let $\sigma=\{R,S\}$, with $ar(R)=2$ and $ar(S)=1$  be a semi-algebraic database schema. Then the
structure ${\cal{D}}$ given by $$(\R,R^{\cal{D}}=\{(x_1,x_2)\in\R^2\mid x_1^2+x_2^2<1\},
S^{\cal{D}}=\{ x \in\R \mid 0\leq x\leq 1 \})$$ is an example of a semi-algebraic database over
$\sigma$ that contains the open unit disk and the closed unit interval.\qed
\end{example}

\begin{definition}\rm
Let $\sigma$ be a $n$-dimensional semi-algebraic database schema. The language
\fom{\sigPolyNoBr}{$\sigma$} (or \foms{\sigPolyNoBr} , if $\sigma$ is clear from the context),
first-order logic over the real numbers with polynomial constraints, is the first-order language
with variables that are assumed to range over real numbers, where the atomic formulas are either of
the form $p(x_1, x_2, \ldots, x_n) > 0$, with $p$ a polynomial with integer coefficients in the
real variables $x_1, x_2, \ldots, x_n$, or the relation names from $\sigma$ applied to real terms.
Atomic formulas are composed using the operations $\mathrel{\land}$, $\vee$ and $\neg$ and the
quantifiers $\forall$ and $\exists$.\qed
\end{definition}

\begin{example}\rm
Consider the semi-algebraic database from Example~\ref{sadb}. The expression $$R(x,y)
\mathrel{\land} y
> 0$$ is a \fom{\sigPolyNoBr}{$\{R, S\}$}-formula selecting the part of the open unit disk that lies strictly above the $x$-axis.\qed
\end{example}

We restrict all further definitions and results to dimension $n = 2$, as this is the dimension we
will be working with in the rest of this text, although they were originally proved to hold for
arbitrary $n$, $n \geq 2$.

Now we give the definition of a geometric database, a special type of constraint database that contains a possibly infinite number of points.

\begin{definition}\rm
Let $\sigma$ be a geometric database schema. A {\em geometric database} over $\sigma$ in $\R^2$ is
a structure $\cal{D}$ over $\sigma$ with domain $\R^2$ such that, for each relation name $R$ of
$\sigma$, the associated relation  $R^{\cal{D}}$ in ${\cal D}$ is semi-algebraic. \qed
\end{definition}

A geometric database $\cal{D}$ over $\sigma$ in $\R^2$ can be viewed naturally as a semi­algebraic
database $\overline{\cal{D}}$ over the schema $\overline{\sigma}$, which has, for each relation
name $R$ of $\sigma$, a relation name $\overline{R}$ with arity $2k$, where $k$ is the arity of $R$
in $\sigma$. For each relation name $R$, of arity $k$, $\overline{R}^{\overline{\cal{D}}}$ is
obtained from $R^{\cal{D}}$ by applying the {\em canonical bijection}\footnote{The canonical
bijection between $(\R^2)^k$ and $\R^{2k}$ associates with each $k$-tuple $(\bf{x}_1, \ldots,
\bf{x}_k)$ of $(\R^2)^k$ the $2k$-tuple $(x_1^1, x_1^2, \ldots, x_k^1, x_k^2)$, where for $1\leq
i\leq k$ ${\bf{x}_i} = (x_i^1, x_i^2)$.} between $(\R^2)^k$ and $\R^{2k}$.

\begin{definition}\rm
Let $\sigma$ be a geometric database schema. A {\em $k$-ary geometric query Q} over $\sigma$ in
$\R^2$ is a partial computable function on the set of geometric databases over $\sigma$.
Furthermore, for each geometric database $\cal{D}$ over $\sigma$ on which $Q$ is defined,
$Q(\cal{D})$ is a geometric relation of arity $k$. \qed
\end{definition}

Queries that are invariant under some transformation group $G$ of $\R^2$, are also called {\em
$G$-generic}~\cite{pvv-pods}. We define this next:

\begin{definition}\rm\label{genericity}
Let $\sigma$ be a geometric database schema and Q a geometric query over $\sigma$ in $\R^2$. Let
$G$ be a group of transformations of $\R^2$. Then $Q$ is called {\em $G$-generic} if, for any two
geometric databases $\cal{D}$ and $\cal{D'}$ over $\sigma$ in $\R^2$ for which $\cal{D'} = $
$g$$(\cal{D})$, for some $g \in G$, we have that $Q(\cal{D'}) = $ $g$$(Q(\cal{D}))$. \qed
\end{definition}

In the remainder of this text, we will only focus on the group $G$ of {\em affinities}. The
affinities of $\R^2$ form the group of linear transformations having a regular matrix, i.e., their
matrix has a determinant different from zero. Affinities of the plane have the following form:
$$\left(
\begin{array}{@{}c@{}}
x \\  y
 \end{array}
 \right) \mapsto \left(
\begin{array}{@{}cc@{}}
a & b \\ c & d\\  \end{array}
 \right) \left(
\begin{array}{@{}c@{}}
x\\  y  \end{array}
 \right) + \left(
\begin{array}{@{}c@{}}
e \\  f
 \end{array}
 \right), $$ where $ad-bc$ is different from zero.

We now give the definition of the first-order point logic \fob, a first-order language where the
variables are not interpreted as real numbers, as in \foms{\sigPolyNoBr} , but as $2$-dimensional
points.

We first introduce the point predicate  {\bf Between}.

\begin{definition}\rm
Let $p = (p _x, p_y)$, $q = (q_x, q_y)$ and $r = (r_x, r_y)$ be points in the plane. The expression
$\predi{Between}(p, q, r)$ is true if and only if either $q$ lies on the closed line segment between $p$ and
$r$ or $p$ and/or $q$ and/or $r$ coincide.\qed
\end{definition}

\begin{figure}[h]
\centerline{ \psfig{file=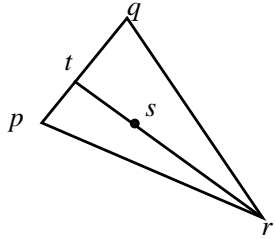,width=100pt}} \caption{The predicate {\bf InTriangle} can
be expressed using {\bf Between}.} \label{figintriangle}
\end{figure}

In Figure~\ref{figintriangle}, $\predi{Between}(p,t,q)$, $\predi{Between}(p,p,q)$ and
$\predi{Between}(t,s,r)$ are true. On the other hand,  but $\predi{Between}(t,q,p)$ and $\predi{Between}(p,q,r)$ are
not true.

\begin{definition}\rm
Let $\sigma$ be a $2$-dimensional geometric database schema. The first-order point language over
$\sigma$ and $\{\bf Between\}$, denoted by \fom{$\{\betws\}$}{$\sigma$} (or, if $\sigma$ is clear
from the context, by \fob), is a first-order language with variables that range over points in
$\R^2$, (denoted $\hat{p}, \hat{q}, \ldots$), where the atomic formulas are equality constraints on
point variables,
 the predicate {\bf Between} applied to point variables, and the relation names from $\sigma$ applied to point
 variables.\qed
\end{definition}

A \fob-formula $\varphi(\hat{p_1}, \hat{p_2}, \ldots, \hat{p_l})$ over the relation names of
$\sigma$ and the predicate {\bf Between} defines on each geometric database $\cal{D}$ over $\sigma$
a subset $\varphi(\cal{D})$ of $(\R^2)^l$ in the standard manner.

Gyssens, Van den Bussche and Van Gucht have shown that the language \fob\ expresses exactly
all affine-generic geometric queries expressible in  \foms{\sigPolyNoBr}.

\section{Notations}\label{sec-notations}

In this Section, we introduce triangle variables and constants. We work in $\Rn{2}$.  Spatial
triangle variables will be denoted $\tr{1}, \tr{2}, \ldots$. Constants containing such triples of
points will be denoted $\trc{1}, \trc{2}, \ldots$, or $\trc{\vect{a}\vect{b}\vect{c}}$ when we want
to emphasize the relationship between a triangle and its corner points $\vect{a}$, $\vect{b}$,
$\vect{c}\in \Rn{2}$. We remark that triangles can be modelled as triples of points in $\Rn{2}$.
Occasionally, we will need to refer to the area of a triangle. The area of a triangle $\trc{}$ will
be abbreviated $A(\trc{})$.

We also introduce \st triangles, which can be modelled as triples of moving points in $\Rn{2}$.
Variables referring to \st triangles are distinguished from spatial triangle variables by a
superscript: $\trST{1}, \trST{2}, \ldots$. The same holds for constants, which are denoted
$\trcST{1}, \trcST{2}, \trcST{pqr}\ldots$. We will also define triangle databases. For the spatial
and \st case respectively, we will use the symbols $\structTr$ and $\structTrST$ to indicate
triangle database instances.

The names of (\st) triangle relations and database schemas containing such relation names will be
recognizable by their hat: $\Rtr$, $\sigmatr$ and $\RtrST$, $\sigmatrST$, respectively. Spatial and
spatio-temporal point relation names and schemas are denoted $\Rpt$ and $\sigmapt$, $\RptST$ and
$\sigmaptST$, respectively.

\section{Definitions}\label{sec-definitions}

We start with the definition of a {\em triangle database}, \ie, a database that contains a
(possibly infinite) collection of triangles. We define both spatial triangle databases and \st
triangle databases. We model triangles by triples of points of $\Rn{2}$, \ie, by elements of
$\Rnm{2}{3}$. Moving or changing (\ie, \st) triangles are modelled by sets of triples of \cotemp
points in $\RnR{2}$,\ie, by sets of elements of $(\Rn{2} \times\{\tau_0\})^3$, for some $\tau_0 \in
\R$. Triangles can degenerate, \ie, corner points are allowed to coincide. For the remainder of
this text, the term triangle refers to a triple of points. We refer to the set of points that is
represented by a triangle as the \emph{drawing} of that triangle.

\begin{definition}[\textbf{Drawing of a triangle}]\label{def-tr-drawing}\rm

\par\noindent$\bullet$ Let $\trc{} = (\vect{a}_1, \vect{a}_2, \vect{a}_3) \in \Rnm{2}{3}$ be a spatial triangle. The {\em drawing}
of $\trc{}$ is the subset of $\Rn{2}$ that is the convex closure of the points $\vect{a}_1$,
$\vect{a}_2$ and $\vect{a}_3$.

\par\noindent$\bullet$ Let $\trcST{} = (p_1, p_2, p_3) \in \RnRm{2}{3}$ be a \st triangle. The {\em drawing}
of $\trcST{}$ is the subset of \cotemp points of $\RnR{2}$ that is the convex closure of the points
$p_1$, $p_2$ and $p_3$.\qed
\end{definition}

We now introduce four bijections.

\begin{itemize}
     \item $can: \Rnm{n}{k} \rightarrow \Rn{nk}$ maps tuples $(\vect{a}_1, \ldots,
\vect{a}_k)$ to $(a_{1,1}, \ldots, a_{1,n}, \ab \ldots,\ab  a_{k,1}, \ab \ldots, \ab a_{k,n})$,
where for $1 \leq i \leq k$ and $1 \leq j \leq n$, $a_{i,j}$ denotes the $j$th real coordinate of
$\vect{a}_i$;
    \item $can_{ST}: (\R^n \times\R)^k\rightarrow \R^{(n+1)\times k}$ maps tuples $((\vect{a}_1, \tau_1), \ldots,
(\vect{a}_k,\tau_k))$ to $(a_{1,1}, \ldots, a_{1,n},\ab \tau_1, \ab \ldots,\ab  a_{k,1}, \ab
\ldots, \ab a_{k,n}, \ab \tau_k)$, where for $1 \leq i \leq k$ and $1 \leq j \leq n$, $a_{i,j}$
denotes the $j$th real coordinate of $\vect{a}_i$;
    \item $can_{tr}: \Rnmk{2}{3}{k} \rightarrow \Rnm{2}{3k}$ maps
$k$-tuples of triangles to $(3k)$ tuples of points in $\Rn{2}$; and
    \item $can_{trST}: \RnRmk{2}{3}{k}
\rightarrow \RnRm{2}{3k}$ maps $k$-tuples of \st triangles to $(3k)$-tuples of points in $\RnR{2}$.
\end{itemize}

\begin{definition}[\textbf{Triangle relations and databases}]\label{def-triangle-base}\rm 
\par\noindent A \emph{(triangle) database sche\-ma} $\sigmatr$ is a finite set of
relation names, where each relation name $\Rtr$ has a natural number $ar(\Rtr)$, called its arity,
associated to it.
\medskip
\par\noindent $\bullet$ A subset ${\cal C}$ of $\Rnmk{2}{3}{k}$ is a
\emph{spatial triangle
   relation of arity $k$} if
\begin{enumerate}[(i)]
    \item its image under the canonical
    bijection $can \circ can_{tr}: \Rnmk{2}{3}{k} \rightarrow \Rn{6k}$ is a semi-algebraic relation of arity
    $6k$, and
    \item for each element $c= ((\vect{a}_{1,1},\vect{a}_{1,2}, \vect{a}_{1,3} ), \ab(\vect{a}_{2,1}, \vect{a}_{2,2}, \vect{a}_{2,3}), \ab\ldots, \ab(\vect{a}_{k,1}, \vect{a}_{k,2}, \vect{a}_{k,3})) \ab\in {\cal
    C}$, also the elements $((\vect{a}_{1,j_{1,1}},\vect{a}_{1,j_{1,2}}, \vect{a}_{1,j_{1,3}} ), \ab(\vect{a}_{2,j_{2,1}}, \vect{a}_{2,j_{2,2}}, \vect{a}_{2,j_{2,3}}), \ab\ldots, \ab(\vect{a}_{k,j_{k,1}}, \ab\vect{a}_{k,j_{k,2}}, \ab\vect{a}_{k,j_{k,3}}))$ are in ${\cal C}$, where $\sigma_i(1,2,3) = (j_{i,1},j_{i,2},j_{i,3})$ with $1 \leq i \leq k$ and  $\sigma_i  \in {\cal S}_3$ where ${\cal S}_3$  is
    the set of all permutations of $\{1,2,3\}$.
\end{enumerate}

Let $\sigmatr$ be a triangle database schema. A {\em spatial triangle database} over $\sigmatr$ in
$\Rnm{2}{3}$ is a structure $\structTr$ over $\sigmatr$ with domain $\Rnm{2}{3}$ such that, for
each relation name $\Rtr$ of $\sigmatr$, the associated triangle relation ${\Rtr}^{\structTr}$ in
$\structTr$ is a spatial triangle relation of arity $ar(\Rtr)$.

 \medskip
 \par\noindent$\bullet$
A subset ${\cal C}$ of $\RnRmk{2}{3}{k}$ is a \emph{spatio-temporal triangle
    relation of arity $k$} if
\begin{enumerate}[(i)]
    \item its image under the canonical
    bijection $can_{trST}\circ can_{ST}: \RnRmk{2}{3}{k} \rightarrow \Rn{9k}$ is a semi-algebraic relation of arity
    $9k$, and
    \item for each element $c= ((p_{1,1},p_{1,2}, p_{1,3} ), \ab(p_{2,1}, p_{2,2}, p_{2,3}), \ab\ldots, \ab(p_{k,1}, p_{k,2}, p_{k,3})) \ab\in {\cal
    C}$, also $((p_{1,j_{1,1}}, \ab p_{1,j_{1,2}}, \ab p_{1,j_{1,3}} ), \ab(p_{2,j_{2,1}}, \ab p_{2,j_{2,2}}, \ab p_{2,j_{2,3}}), \ab\ldots, \ab(p_{k,j_{k,1}}, \ab p_{k,j_{k,2}}, \ab p_{k,j_{k,3}}))$ are in ${\cal C}$, where $\sigma_i(1,2,3) = (j_{i,1},j_{i,2},j_{i,3}) (1 \leq i \leq k; \sigma_i  \in {\cal S}_3)$. Here, ${\cal S}_3$  is
    the set of all permutations of $\{1,2,3\}$.
\end{enumerate}

    Let $\sigmatrST$ be a triangle database schema. A {\em \st triangle database} over $\sigmatrST$ is a
structure $\structTrST$ over $\sigmatrST$ with domain $\RnRm{2}{3}$ such that, for each relation
name $\RtrST$ of $\sigmatrST$, the associated triangle relation ${\hat{R}}^{{st}\structTrST}$ in
$\structTrST$ is a \st triangle relation of arity $ar(\RtrST)$.\qed
\end{definition}

\medskip

We want to remark two things about the definition of triangle relations (as given in
Definition~\ref{def-triangle-base}), one about the items (i) and one about the items (ii) of the
definition of triangle relations. They are discussed in Remark~\ref{remark-triangle-can} below and
Remark~\ref{remark-triangle-perm}, which is   postponed until after the definition of triangle
database queries.

\begin{remark}\label{remark-triangle-can}\rm
 A triangle database $\structTr$ over $\sigmatr$ in $\Rnm{2}{3}$
can be viewed naturally as a geometric database $\structPt$ over the schema $\sigmapt$, which has,
for each relation name $\Rtr$ of $\sigmatr$, a relation name $\Rpt$ with arity $3\times ar(\Rtr)$.
For each relation name $\Rtr$, of arity $k$, ${\Rpt}^{\structPt}$ is obtained from
${\Rtr}^{\structTr}$ by applying the canonical bijection $can_{tr}: \Rnmk{2}{3}{k} \rightarrow
\Rnm{2}{3k}$. Analogously, a spatio-temporal triangle database $\structTrST$ over $\sigmatrST$ can
be viewed naturally as a \st database $\structST$ over the schema $\sigmaptST$, which has, for each
relation name $\RtrST$ of $\sigmatrST$, a relation name $\RptST$ with arity $3\times ar(\RtrST)$.
For each relation name $\RtrST$, of arity $k$, ${\Rpt}^{{st}\structST}$ is obtained from
${\Rtr}^{{st}\structTrST}$ by applying the canonical bijection $can_{trST}: \RnRmk{2}{3}{k}
\rightarrow \RnRm{2}{3k}$.\qed
\end{remark}

\begin{example}\rm\label{ex-def-tbase}\rm 
It follows from the definition of triangle relations that they can be finitely represented by
polynomial constraints on the coordinates of the corner points of the triangles they contain.

For example, the unary spatial triangle relation containing all triangles with one corner point on
the $x$-axis, one on the $y$-axis and a third corner point on the diagonal $y = x$, can be finitely
represented as follows:

$$\displaylines{\qquad \{(\vect{z}_1, \vect{z}_2, \vect{z}_3) = ((z_{1,x}, z_{1,y}), (z_{2,x}, z_{2,y}), (z_{3,x}, z_{3,y})) \in \Rnm{2}{3}~\mid\splits
 (z_{1,x} = 0 \land z_{2,y} = 0 \land z_{3,x} = z_{3,y}) \lor (z_{1,x} = 0 \land z_{3,y} = 0 \land z_{2,x} = z_{2,y})\splits
  \lor (z_{2,x} = 0 \land z_{1,y} = 0 \land z_{3,x} = z_{3,y}) \lor (z_{2,x} = 0 \land z_{3,y} = 0 \land z_{1,x} = z_{1,y})\splits
   \lor (z_{3,x} = 0 \land z_{2,y} = 0 \land z_{1,x} = z_{1,y}) \lor (z_{3,x} = 0 \land z_{1,y} = 0 \land z_{2,x} = z_{2,y})\}.\qquad}$$

\begin{figure}
  \centerline{\includegraphics[width=150pt]{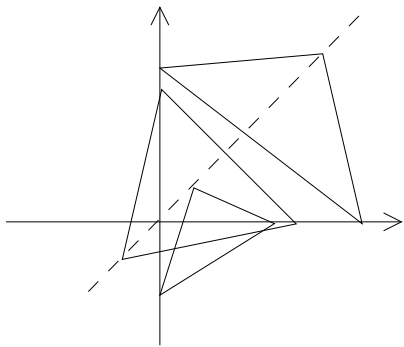}}
  \caption{Some elements of the relation represented in Example~\ref{ex-def-tbase}.}\label{fig-extbase}
\end{figure}

Figure~\ref{fig-extbase} gives some elements of this relation. Each triangle that is drawn is
stored three times in the relation.
\qed\end{example}

\begin{remark}\label{remark-datastructure}\rm
For the remainder of this text we assume that databases are finitely encoded by systems of
polynomial equations and that a specific data structure is fixed (possible data structures are
dense or sparse representations of polynomials). The specific choice of data structure is not
relevant to the topic of this text, but we assume that one is fixed. When we talk about computable
queries later on, we mean Turing computable with respect to the chosen encoding and data
structures.\qed
\end{remark}

We also remark the following.
\begin{remark}\label{remark-thematic}\rm 
 The data model and the query languages presented in this text can be
extended straightforwardly to the situation where \st relations are accompanied by classical
thematic information. However, because the problem that is discussed here is captured by this
simplified model, we stick to it for reasons of simplicity of exposition. \qed\end{remark}

We now define spatial and \st triangle database queries.

\begin{definition}[\textbf{Triangle database queries}]\label{def-triangle-query}\rm

$\bullet$ Let $\sigmatr$ be a triangle database schema and let us consider input spatial triangle
databases over $\sigmatr$. A {\em $k$-ary spatial triangle database query $Q$} over $\sigmatr$  is
a computable partial mapping (in the sense of Remark~\ref{remark-datastructure}) from the set of
spatial triangle databases over $\sigmatr$ to the set of $k$-ary spatial triangle relations.

$\bullet$ Let $\sigmatrST$ be a database schema and let us consider input \st triangle databases
over $\sigmatrST$. A {\em $k$-ary \st triangle database query $Q$} over $\sigmatrST$ is a
computable partial mapping (in the sense of Remark~\ref{remark-datastructure}) from the set of \st
triangle databases over $\sigmatrST$ to the set of $k$-ary \st triangle relations.\qed
\end{definition}

\begin{remark}\label{remark-triangle-perm}\rm
In the (ii)-items of the definition of triangle relations, we require that, if a triangle $\trc{}$
is involved in a relation, that also all other triangles with the same drawing are stored in that
relation. The reason for this is that we do not want the triangle queries to be dependent of the
actual order and orientation used when enumerating the corner points of a triangle. When
emphasizing property (ii) of a relation, we will call it \emph{consistency} and talk about
\emph{consistent triangle relations}. Also, a database is said to be consistent, if all its
relations are consistent. \qed\end{remark}

We illustrate the consistency property with some examples:

\begin{example}\rm\label{ex-consistency}\rm
Let $\sigmatr = \{\Rtr\}$ be a database schema. First, we list some queries over $\sigmatr$ that
are not consistent:
\medskip
\par\noindent $\bullet$ {\em Q$_6$: Give all triangles in $\Rtr$ for which their first and second corner points coincide.}
\medskip
\par\noindent $\bullet$ {\em Q$_7$: Give all triangles for which the segment defined by their first and second corner point is a boundary segment of one of the triangles in $\Rtr$.}
\medskip
\par\noindent Now some consistent queries follow:
\medskip
\par\noindent $\bullet$  {\em Q$_8$: Give all triangles in $\Rtr$ that are degenerated into a line segment.}
\medskip
\par\noindent $\bullet$ {\em Q$_9$: Give all triangles that share a boundary segment with some triangle in $\Rtr$.}

It is clear that the inconsistent queries are rather artificial. When a user specifies the
triangles that should be in the result of a query, she intuitively thinks of the drawings of those
triangles. The order of the corner points used in the construction of those triangles should not be
important. \qed\end{example}

\begin{remark}\label{remark-canon}
\rm A \st database $\stb{}$ over $\sigmaST$ can be viewed in a natural way as a constraint database
$D$ over the constraint schema $\sigma$, which has for each relation name $\RST$ of $\sigmaST$, a
relation name $R$ of arity $(n+1)\times ar(\RST)$. For each relation name $\RST$, $R^{D}$ is
obtained from ${\RST}^{\stb{}}$ by applying the canonical bijection $can_{ST}: (\R^n \times
\R)^{ar(R)}\rightarrow \R^{(n+1)\times ar(R)}$. We will use the notation introduced here,
throughout this text. \qed\end{remark}

 Analogously, spatial and \st triangle database
queries can be seen as constraint queries, and as spatial and \st (point) database queries. We
prefer the latter view, as we already developed affine-generic \st point languages in
\cite{ghk-01}, and there already exist affine-generic spatial point languages\cite{gvv-jcss}. We
define the equivalence between triangle queries and point queries formally:

\begin{definition}[\textbf{Equivalence of point queries and triangle queries}]\label{def-triangle-query-equiv}\rm

\par\noindent$\bullet$ Let $\sigmatr$ be a triangle database schema and let us
consider input spatial triangle databases over $\sigmatr$. Let $\sigmapt$ be the corresponding
spatial point database schema (see Remark~\ref{remark-triangle-can}). Let $\qtr$ be a $k$-ary
spatial triangle database query over $\sigmatr$ and let $\qpt$ be a $(3k)$-ary spatial (point)
database query over $\sigmapt$. We say that \emph{$\qtr$ and $\qpt$ are equivalent}, denoted $\qtr
\equiv_{\vartriangle}\qpt$ if for every database $\structTr$ over $\sigmatr$ we have
$$can_{tr}(\qtr(\structTr  )) = \qpt(can_{tr}(\structTr)).$$

\par\noindent$\bullet$ Let $\sigmatrST$ be a triangle database schema and let us consider input \st triangle databases over
$\sigmatrST$. Let $\sigmaptST$ be the corresponding \st point database schema (see
Remark~\ref{remark-triangle-can}). Let $\qtr$ be a $k$-ary \st triangle database query over
$\sigmatrST$ and let $\qpt$ be a $(3k)$-ary \st (point) database query over $\sigmaptST$. We say
that \emph{$\qtr$ and $\qpt$ are equivalent}, denoted $\qtr \equiv_{\vartriangle}\qpt$,  if for
every database $\structTrST$ over $\sigmatrST$ we have
$$can_{trST}(\qtr(\structTrST)) = \qpt(can_{trST}(\structTrST)).$$\qed
\end{definition}

Since we have defined equivalence between triangle database queries and point database queries
earlier, we can now discuss how the point languages \foms{$\{\betws\{$} and \fobST can be used to
query triangle databases. We have to keep in mind that only spatial and \st (point) databases can
be considered that are the image under the bijections $can_{tr}$ and $can_{trST}$ of spatial and
\st triangle databases.

\begin{definition}[\textbf{\fob\ as a triangle query language}]\label{def-triangle-fob-is-poss-lang}\rm

\par\noindent$\bullet$ Let $\sigmatr = \{\Rtr_1, \ab \Rtr_2, \ab \ldots, \ab \Rtr_m\}$ be a
spatial triangle database
    schema. Let $\Rpt_i$ be the corresponding spatial point relation names of
    arity $3\times ar(\Rtr_i)$, for $i = 1 \ldots m$, and let $\sigmapt$ be the spatial database schema $\{\Rpt_1, \Rpt_2, \ldots, \Rpt_m\}$.

    Let $\varphi(\vect{x}_{1,1}, \vect{x}_{1,2}, \vect{x}_{1,3}, \vect{x}_{2,1}, \vect{x}_{2,2}, \vect{x}_{2,3}, \ldots, \vect{x}_{k,1}, \vect{x}_{k,2},
    \vect{x}_{k,3})$ be a \fob-formula expressing a spatial $(3k)$-ary query $\qpt$ which is equivalent to a $k$-ary spatial triangle query
    $\qtr$. For each input spatial triangle database $\structTr$ over $\sigmatr$, $\qtr(\structTr)$
    is defined as the set of points $(\vect{a}_{1,1}, \vect{a}_{1,2}, \vect{a}_{1,3},\ab \vect{a}_{2,1}, \vect{a}_{2,2}, \vect{a}_{2,3}, \ldots, \ab \vect{a}_{k,1}, \vect{a}_{k,2},
    \vect{a}_{k,3})$ in $\Rnm{6}{k}$ such that
    $$\displaylines{\qquad (\Rn{2}, =, \betws,{\Rpt_1}^{\structPt}, {\Rpt_2}^{\structPt}, \ldots, {\Rpt_m}^{\structPt}) \models \splits \varphi[\vect{a}_{1,1}, \vect{a}_{1,2}, \vect{a}_{1,3}, \vect{a}_{2,1}, \vect{a}_{2,2}, \vect{a}_{2,3}, \ldots, \vect{a}_{k,1}, \vect{a}_{k,2},
    \vect{a}_{k,3}].\qquad}$$ Here, $\structPt$ is the image of $\structTr$ under the canonical
    bijection $can_{tr}$.

\par\noindent$\bullet$ Let $\sigmatrST = \{\RtrST_1, \RtrST_2, \ldots, \RtrST_m\}$ be a \st triangle database
    schema. Let $\RptST_i (1 \leq i \leq m)$ be the corresponding \st point relation names of
    arity $3\times ar(\RtrST_i)$ and let $\sigmaptST$ be the \st database schema $\{\RptST_1, \RptST_2, \ldots, \RptST_m\}$.

    Let $\varphi(u_{1,1}, u_{1,2}, u_{1,3}, u_{2,1}, u_{2,2}, u_{2,3}, \ldots, u_{k,1}, u_{k,2},
    u_{k,3})$ be a \fob-for\-mu\-la, expressing a \st $(3k)$-ary query $\qpt$ which is equivalent to a $k$-ary spatial triangle query
    $\qtr$. For each input \st triangle database $\structTrST$ over $\sigmatrST$, $\qtr(\structTrST)$
    is defined as the set of points $(p_{1,1}, p_{1,2},p_{1,3}, \ab p_{2,1}, p_{2,2}, p_{2,3}, \ldots, \ab p_{k,1}, \ab p_{k,2},
    \ab p_{k,3})$ of $\Rnm{9}{k}$ such that
    $$\displaylines{\qquad (\RnR{2}, =, \betwsCotemp,\befores, \eqcrsts,{\Rpt_1}^{{st}\structTrST}, {\Rpt_2}^{{st}\structTrST}, \ldots, {\Rpt_m}^{{st}\structTrST}) \models \splits \varphi[p_{1,1}, p_{1,2},p_{1,3},p_{2,1}, p_{2,2}, p_{2,3}, \ldots, p_{k,1},p_{k,2},
    p_{k,3}].\qquad}$$ Here, $\structST$ is the image of $\structTrST$ under the canonical
    bijection $can_{trST}$.\qed
\end{definition}

The languages \fom{$\{\betws\}$}{$\sigmapt$} and \fobgenSTsigma were designed to formulate queries
on spatial and \st point databases over some input schema $\sigmapt$, resp. $\sigmaptST$. Using
those languages to query triangle databases, involves expressing relations between the point sets
that compose the triangles. This is a rather indirect way of expressing triangle relations. In the
spirit of \cite{ghk-01}, we now construct affine-generic query languages based on triangle
variables. As they directly express relations between the triangles, this results in a more
intuitive way of querying spatial and \st triangle databases. We define triangle-based logics next.
Afterwards, we propose a specific spatial triangle logic in Section~\ref{sec-tr-logic-spatial}, and
a \st triangle logic in Section~\ref{sec-tr-logic-st}.

\begin{definition}[\textbf{Triangle logics}]\label{def-triangle-logic}\rm
\par\noindent$\bullet$ Let $\sigmatr = \{\Rtr_1, \Rtr_2, \ldots, \Rtr_m\}$ be a
triangle database schema and let ${\bf \Delta}$ be a set of predicates of a certain arity over
triangles in $\Rn{2}$. The first-order logic over $\sigmatr$ and ${\bf \Delta}$, denoted by
\fom{${\bf \Delta}$}{$\sigmatr$}, can be used as a spatial triangle query language when variables
are interpreted to range over triangles in $\Rn{2}$. The atomic formulas in \fom{${\bf
\Delta}$}{$\sigmatr$} are equality constraints on triangle variables, the predicates of ${\bf
\Delta}$, and the relation names $\Rtr_1, \Rtr_2, \ldots, \Rtr_m$ from $\sigmatr$, applied to
triangle variables.

\par\noindent$\bullet$ Let $\sigmatrST = \{\RtrST_1, \RtrST_2, \ldots, \RtrST_m\}$ be a
database schema and let ${\bf \Delta}$ be a set of predicates of a certain arity over \st triangles
in $\RnR{2}$. The first-order logic over $\sigmatrST$ and ${\bf \Delta}$, denoted by \fom{${\bf
\Delta}$}{$\sigmatrST$}, can be used as a \st triangle query language when variables are
interpreted to range over \st triangles in $\RnR{2}$. The atomic formulas in \fom{${\bf
\Delta}$}{$\sigmatrST$} are equality constraints on \st triangle variables, the predicates of ${\bf
\Delta}$, and the relation names $\RtrST_1, \RtrST_2, \ldots, \RtrST_m$ from $\sigmatrST$, applied
to \st triangle variables.\qed
\end{definition}

A  \fom{${\bf \Delta}$}{$\sigmatr$}-formula $\varphi(\tr{1}, \tr{2}, \ldots, \tr{k})$ (resp.,
\fom{${\bf \Delta}$}{$\sigmatrST$}-formula $\varphi(\trST{1}, \trST{2}, \ldots, \ab \trST{k})$) defines
for each spatial (resp., \st) database $\structTr$ (resp., $\structTrST$) over $\sigmatr$ (resp.
$\sigmatrST$) a subset $\varphi(\structTr)$ (resp., $\varphi(\structTrST)$) of $\Rnmk{2}{3}{k}$
(resp., $\RnRmk{2}{3}{k}$) defined as
$$\displaylines{\qquad \{(\trc{1}, \trc{2}, \ldots, \trc{k}) \in \Rnm{2}{3k}\mid\splits(\Rn{2}, {\bf \Delta}^{\Rn{2}}, {\Rtr_1}^{\structTr}, {\Rtr_2}^{\structTr}, \ldots, {\Rtr_m}^{\structTr}) \models \varphi[\trc{1}, \trc{2}, \ldots, \trc{k}]~\},\qquad}$$
respectively,
$$\displaylines{\qquad \{(\trcST{1}, \trcST{2}, \ldots, \trcST{k}) \in \RnRm{2}{3k}\mid\splits (\RnR{2}, {\bf \Delta}^{\RnR{2}}, {\Rtr_1}^{{st}\structTrST}, {\Rtr_2}^{{st}\structTrST}, \ldots, {\Rtr_m}^{{st}\structTrST}) \models \varphi[\trcST{1}, \trcST{2}, \ldots, \trcST{k}]~\}.\qquad}$$

\begin{remark}\rm 
We use the symbol $\mathrel{=}_{\vartriangle}$ to indicate equality of triangle variables, as
opposed to equality of point variables. If it is clear from the context of a formula which type of
variables is used, we will omit the index.\qed
\end{remark}

In Section~\ref{sec-tr-logic-spatial} (resp., Section~\ref{sec-tr-logic-st}), we will develop
languages that have the same expressive power as  \fob\  and \fobST on spatial triangle databases
and on \st triangle databases, respectively.  We will prove this by showing both soundness and
completeness of those triangle languages with respect to \fob\ and \fobST.

The concepts of soundness and completeness are introduced as follows:
\begin{definition}[\textbf{Soundness and completeness}]\label{def-sound-complete}\rm
  $\bullet$ A query language ${\cal L}$ is said to be
\emph{sound} for the ${\cal G}$-generic \fom{\sigPolyNoBr}{$\sigma$}-queries on spatial (resp.,
\st) databases, if formulas in ${\cal L}$ only express ${\cal G}_{st}$-generic
\fom{\sigPolyNoBr}{$\sigma$}-queries on spatial (resp., \st) databases.

$\bullet$  A query language ${\cal L}$ is said to be
 \emph{complete} for the $({\cal F}_{st}, {\cal F}_t)$-generic
\fom{\sigPolyNoBr}{$\sigma$}-queries on \st databases,
 if all $({\cal F}_{st}, {\cal F}_t)$-generic \fom{\sigPolyNoBr}{$\sigma$}-queries on \st
databases can be expressed in ${\cal L}$. \qed\end{definition}

\section{Affine-invariant Spatial Triangle Queries}\label{sec-tr-logic-spatial}

In this section, we propose a  spatial triangle logic that captures exactly the class of
first-order affine-generic queries on spatial triangle databases. First, we remark the following:

\medskip

\begin{remark}\label{rem-drawing}\rm We defined a triangle database as a special type of geometric database. Accordingly, we
take the affine image of a triangle for affinities of $\Rn{2}$, and not of $\Rn{6}$. This
corresponds to our intuition. One triangle is an affine image of another triangle, if the drawing
of the first one is the affine image of the drawing of the second one. Hence, the affine image of a
triangle with corner points $\vect{x}_1$, $\vect{x}_2$ and $\vect{x}_3$ under some affinity
$\alpha$ of the plane, is the triangle with corner points $\alpha(\vect{x}_1)$,
$\alpha(\vect{x}_2)$ and $\alpha(\vect{x}_3)$.\qed
\end{remark}

We introduce one binary triangle predicate, \ie, $\pos$. Intuitively, when applied to two
triangles, this predicate expresses that the drawing of the first triangle is a subset
$(\subseteq)$ of the drawing of the second triangle. We only consider $\Rnm{2}{3}$ as the
underlying domain. We show that the triangle predicate $\pos$ allows a natural extension to higher
dimensions and other types of objects (instead of triangles).

 We define the predicate $\pos$ and equality on triangles
more precisely:

\begin{definition}[\textbf{The triangle predicate $\pos$}]\label{def-triangle-predi-partof}\rm
Let $\trc{1} = (\vect{a}_{1,1}, \vect{a}_{1,2}, \vect{a}_{1,3})$ and $\trc{2} = (\vect{a}_{2,1},
\vect{a}_{2,2}, \vect{a}_{2,3})$ be two triangles. The binary predicate $\pos$, applied to
$\trc{1}$ and $\trc{2}$ expresses that the convex closure of the three points $\vect{a}_{1,1}$,
$\vect{a}_{1,2}$ and  $\vect{a}_{1,3}$ is a subset of the convex closure of the three points
$\vect{a}_{2,1}$, $\vect{a}_{2,2}$ and  $\vect{a}_{2,3}$.\qed
\end{definition}

\begin{figure}
  \centerline{\includegraphics[width=150pt]{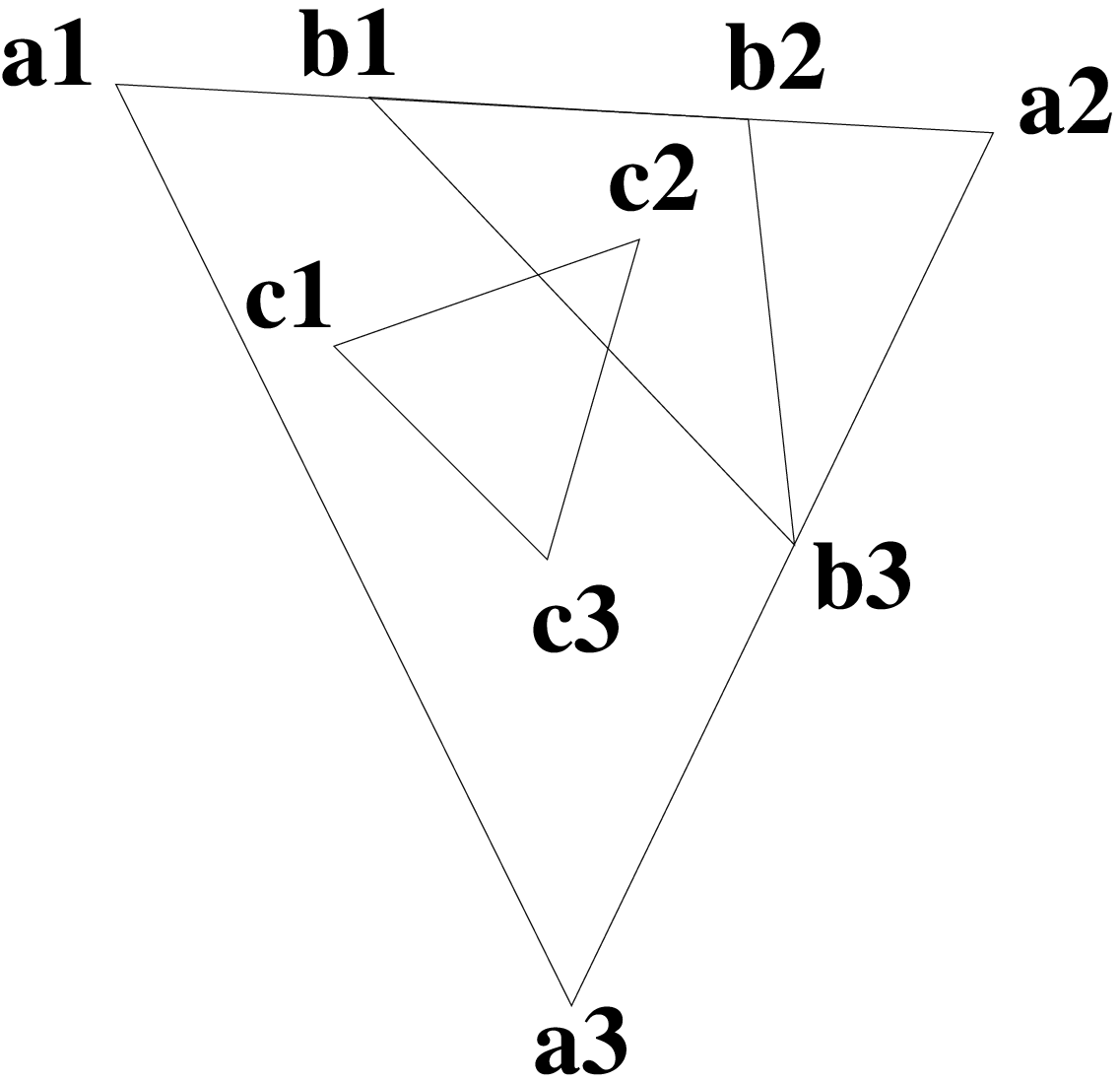}}
  \caption{An illustration of the predicate $\pos$. Let $\trc{1} = (\vect{a}_1, \vect{a}_2,
  \vect{a}_3)$, $\trc{2} = (\vect{b}_1, \vect{b}_2, \vect{b}_3)$ and
  $\trc{3} = (\vect{c}_1, \vect{c}_2, \vect{c}_3)$ The expressions $\po{\trc{2}}{\trc{1}}$ and $\po{\trc{3}}{\trc{1}}$ are true, the expression $\po{\trc{3}}{\trc{2}}$ is not true.}\label{fig-pos-example}
\end{figure}

Figure~\ref{fig-pos-example} illustrates the predicate $\pos$.

We also define triangle-equality, which differs from the standard equality operation.

\begin{definition}[\textbf{Equality of triangles}]\label{def-triangle-predi-eq}\rm  Let $\trc{1}$ and $\trc{2}$ be two triangles. The
expression $\trc{1} =_{\vartriangle} \trc{2}$ is true if and only if both $\po{\trc{1}}{\trc{2}}$ and
$\po{\trc{2}}{\trc{1}}$ are tue.\qed
\end{definition}

Before analyzing the expressiveness of the language \fod, we prove that the \fod-queries are
well-defined on consistent triangle databases. More concretely, given a triangle database schema
$\sigmatr$, we prove that the result of a $k$-ary \fom{$\Delta$}{$\sigmatr$} query on a consistent
input database over $\sigmatr$ is a consistent triangle relation of arity $k$.

\begin{lemma}[\textbf{\fod is well-defined}]\rm\label{lemma-tr-well-defined}
Let $\sigmatr = \{\Rtr_1, \ab\Rtr_2, \ab\ldots, \ab\Rtr_m\}$ be a spatial triangle database
    schema. Let $\structTr$ be a consistent spatial triangle database over $\sigmatr$.
For each \fom{$\Delta$}{$\sigmatr$}-query $\qtr$, $\qtr(\structTr)$ is a consistent triangle
relation.
\end{lemma}
\par\noindent{\bf Proof.} Let $\sigmatr = \{\Rtr_1, \ab\Rtr_2, \ab\ldots, \ab\Rtr_m\}$ be a spatial triangle database
    schema. Let $\structTr$ be a consistent spatial triangle database over $\sigmatr$.

We prove this lemma by induction on the structure of \fom{$\Delta$}{$\sigmatr$}-queries. The atomic
formulas of \fod are equality expressions on triangle variables, expressions of the form
$\predi{PartOf}(\tr{1}, \ab\tr{2})$, and expressions of the form ${\Rtr_i}(\tr{1}, \tr{2},
\ldots,\ab \tr{ar(\Rtr_i)})$, where $\Rtr_i (1 \leq i \leq m)$ is a relation name from $\sigmatr$.
More complex formulas can be constructed using the Boolean operators $\land$, $\lor$ and $\neg$ and
existential quantification.

For the atomic formulas, it is easy to see that, if two triangles $\trc{1}$ and $\trc{2}$ satisfy
the conditions $\trc{1} =_{\vartriangle} \trc{2}$ or $\po{\trc{1}}{\trc{2}}$, that also $\trc{1}'
=_{\vartriangle} \trc{2}'$ respectively $\po{\trc{1}'}{\trc{2}'}$ are true if and only if $\trc{1}
=_{\vartriangle}\trc{1}'$ and $\trc{2}=_{\vartriangle}\trc{2}'$ are true. As we assume the input
database $\structTr$ to be consistent, the atomic formulas of the type $\Rtr_i(\tr{1}, \tr{2},
\ldots, \tr{ar(\Rtr_i)})$, where $(1 \leq i \leq m)$, trivially return consistent triangle
relations.

Now we have to prove that the composed formulas always return consistent triangle relations. Let
$\hat{\varphi}$ and $\hat{\psi}$ be two formulas in \fom{$\Delta$}{$\sigmatr$}, of arity
$k_{\varphi}$ and $k_{\psi}$ respectively, already defining consistent triangle relations. Then,
the formula $(\hat{\varphi} \land \hat{\psi})$ (resp., $(\hat{\varphi} \lor \hat{\psi})$) also
defines a triangle relation. This follows from the fact that the free variables of
$(\hat{\varphi}\land\hat{\psi})$ (resp., $(\hat{\varphi} \lor \hat{\psi})$) are free variables in
$\hat{\varphi}$ or $\hat{\psi}$. The universe of all triangles is trivially consistent. If a
consistent subset is removed from this universe, the remaining part is still consistent. Therefor,
$\neg\hat{\varphi}$ is well-defined. Finally, because consistency is defined argument-wise, the
projection $\exists\trc{1}\,\hat{\varphi}(\trc{1}, \trc{2}, \ldots, \trc{k_{\varphi}})$ is
consistent. \qed

\medskip

After proving that the language \fod is well-defined, we can analyze its expressiveness.

\subsection{Expressiveness of \fod}\label{sec-expressiveness-fod}

We now determine the expressiveness of the language \fod. We prove that it is sound and complete
for the affine-invariant fragment of \fol logic over the reals, on triangle databases. We prove
this by comparing the languages \fod and \foms{$\{\betws\}$}. 

From \cite{gvv-jcss}, we already know
that \fob\ is sound and complete for the affine-invariant fragment of \fol logic over the reals, on
spatial point databases.

The soundness and completeness of  the query language \fod with respect to the language
\foms{$\{\betws\}$} is proved using two separate lemmas (Lemma~\ref{lemma-tr-sound} and
Lemma~\ref{lemma-tr-complete}). In both lemmas, formulas are translated from one language in the
other, by using induction on the structure of \fod and \foms{$\{\betws\}$}-formulas, respectively.
This proof technique will be used several times in this text. Therefor, we explain the first such
proofs in detail. Later on, we will only develop the crucial points in similar proofs.

\begin{lemma}[\textbf{Soundness of} \fod {\bf with respect to} \foms{$\{\betws\}$}]\rm\label{lemma-tr-sound}
Let $\sigmatr = \{\Rtr_1, \ab\Rtr_2, \ab\ldots, \ab\Rtr_m\}$ be a spatial triangle database
    schema. Let $\Rpt_i$ be the corresponding spatial point relation names of
    arity $3\times ar(\Rtr_i)$, for  $(1 \leq i \leq m)$, and let $\sigmapt$ be the spatial database schema $\{\Rpt_1, \Rpt_2, \ldots, \Rpt_m\}$.
    Every \fom{$\Delta$}{$\sigmatr$}-expressible query can be expressed equivalently in \fom{$\{\betws\}$}{$\sigmapt$}.
\end{lemma}
\par\noindent{\bf Proof.}
Let $\sigmatr = \{\Rtr_1, \ab\Rtr_2, \ab\ldots, \ab\Rtr_m\}$ be a spatial triangle database
    schema. Let $\Rpt_i$ be the corresponding spatial point relation names of
    arity $3\times ar(\Rtr_i)$, for  $(1 \leq i \leq m)$, and let $\sigmapt$ be the corresponding spatial database schema $\{\Rpt_1, \Rpt_2, \ldots, \Rpt_m\}$.
    We translate each formula of \fom{$\Delta$}{$\sigmatr$} into an equivalent formula in
\fom{$\{\betws\}$}{$\sigmapt$}. We do this by induction on the structure of
\fom{$\Delta$}{$\sigmatr$}-formulas.

First, we translate the variables of $\hat{\varphi}$. Each triangle variable $\tr{}$ is naturally
translated into three spatial point variables $\vect{x}_{1}$, $\vect{x}_{2}$ and $\vect{x}_{3}$. We
allow one or more of the corner points of a triangle to coincide, so there are no further
restrictions on the variables $\vect{x}_j, 1 \leq j \leq 3$.

The atomic formulas of \fom{$\Delta$}{$\sigmatr$} are equality expressions on triangle variables,
expressions of the form $\predi{PartOf}(\tr{1}, \ab\tr{2})$, and expressions of the form
${\Rtr_i}(\tr{1}, \tr{2}, \ldots,\ab \tr{k})$, where $k = ar(\Rtr_i)$ and $1 \leq i \leq m$. More
complex formulas can be constructed using the Boolean operators $\land$, $\lor$ and $\neg$ and
existential quantification.

\medskip
\par\noindent{\it The translation of atomic formulas.}
\medskip
\par
\begin{figure}
\begin{center}
\begin{picture}(0,0)%
\includegraphics{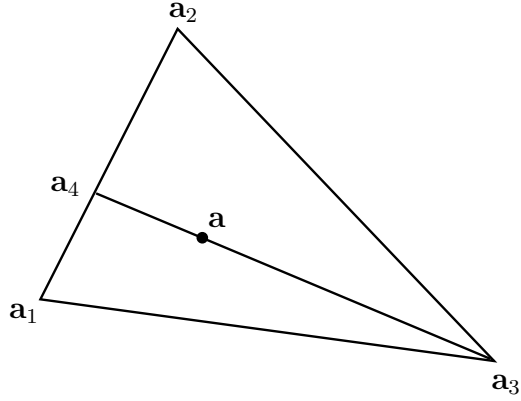}%
\end{picture}%
\setlength{\unitlength}{3947sp}%
\begingroup\makeatletter\ifx\SetFigFont\undefined%
\gdef\SetFigFont#1#2#3#4#5{%
  \reset@font\fontsize{#1}{#2pt}%
  \fontfamily{#3}\fontseries{#4}\fontshape{#5}%
  \selectfont}%
\fi\endgroup%
\begin{picture}(3082,2555)(854,-2141)
\put(1871,260){\makebox(0,0)[lb]{\smash{{\SetFigFont{12}{14.4}{\rmdefault}{\mddefault}{\updefault}{\color[rgb]{0,0,0}$\mathbf{a}_2$}%
}}}}
\put(869,-1631){\makebox(0,0)[lb]{\smash{{\SetFigFont{12}{14.4}{\rmdefault}{\mddefault}{\updefault}{\color[rgb]{0,0,0}$\mathbf{a}_1$}%
}}}}
\put(2118,-1059){\makebox(0,0)[lb]{\smash{{\SetFigFont{12}{14.4}{\rmdefault}{\mddefault}{\updefault}{\color[rgb]{0,0,0}$\mathbf{a}$}%
}}}}
\put(1129,-833){\makebox(0,0)[lb]{\smash{{\SetFigFont{12}{14.4}{\rmdefault}{\mddefault}{\updefault}{\color[rgb]{0,0,0}$\mathbf{a}_4$}%
}}}}
\put(3896,-2075){\makebox(0,0)[lb]{\smash{{\SetFigFont{12}{14.4}{\rmdefault}{\mddefault}{\updefault}{\color[rgb]{0,0,0}$\mathbf{a}_3$}%
}}}}
\end{picture}%
\end{center}
   \caption{An illustration of the predicate $\predi{InTriangle}$. The expression $\predi{InTriangle}(\vect{a}, \vect{a}_1, \vect{a}_2, \vect{a}_3)$ is true because there exists a point $\vect{a}_4$ between $\vect{a}_1$ and $\vect{a}_2$ such that $\vect{a}$ lies between $\vect{a}_4$ and $\vect{a}_3$.}\label{fig-intriangle}
\end{figure}

\noindent We first show that all atomic formulas of \fom{$\Delta$}{$\sigmatr$} can be expressed in the
language \fobsigma.
\begin{enumerate}[(i)]
    \item  The translation of $(\tr{1} \mathrel{=} \tr{2})$, where $\trc{1}$ is translated into $\vect{x}_{1,1}$, $\vect{x}_{1,2}$ and $\vect{x}_{1,3}$ and $\trc{2}$ is translated into $\vect{x}_{2,1}$, $\vect{x}_{2,2}$ and $\vect{x}_{2,3}$, equalsLet $\trcST{1} = (p_{1,1}, p_{1,2}, p_{1,3})$ and $\trcST{2} =
(p_{2,1}, p_{2,2}, p_{2,3})$ be two triangle snapshots. The binary predicate $\pos$, applied to
$\trcST{1}$ and $\trcST{2}$ expresses that $p_{1,1}$, $p_{1,2}$ and $p_{1,3}$ (resp., $p_{2,1}$,
$p_{2,2}$ and $p_{2,3}$) are \cotemp and that the convex closure of the three points $p_{1,1}$,
$p_{1,2}$ and $p_{1,3}$ is a subset of the convex closure of the three points $p_{2,1}$, $p_{2,2}$
and $p_{2,3}$.

$$\displaylines{\qquad \bigvee_{\sigma(1,2,3) = (j_1, j_2, j_3), \sigma \in {\cal S}_3} (\vect{x}_{1,1} \mathrel{=} \vect{x}_{2,{j_1}}\land \vect{x}_{1,2} \mathrel{=} \vect{x}_{2,{j_2}}\land \vect{x}_{1,3} \mathrel{=} \vect{x}_{2,{j_3}}),\qquad}$$
where ${\cal S}_3$ is the set of all permutations of $\{1,2,3\}$.

The correctness of this translation follows trivially from the definition of triangle equality (see
Definition~\ref{def-triangle-predi-eq}).
    \item The translation of $\po{\tr{1}}{\tr{2}}$, where $\trc{1}$ is translated into $\vect{x}_{1,1}$, $\vect{x}_{1,2}$ and $\vect{x}_{1,3}$ and $\trc{2}$ is translated into $\vect{x}_{2,1}$, $\vect{x}_{2,2}$ and $\vect{x}_{2,3}$,
    is $$\displaylines{\qquad\bigwedge_{i=1}^3 \predi{InTriangle}(\vect{x}_{1,i}, \vect{x}_{2,1}, \vect{x}_{2,2},
\vect{x}_{2,3}),\qquad}$$ where the definition of $\predi{InTriangle}$ is:
$$\displaylines{\qquad\predi{InTriangle}(\vect{x}, \vect{x}_1, \vect{x}_2, \vect{x}_3) := \splits\exists\,\vect{x}_4(\betw{\vect{x}_1}{\vect{x}_4}{\vect{x}_2} \mathrel{\land}
\betw{\vect{x}_4}{\vect{x}}{\vect{x}_3}).\qquad}$$ Figure~\ref{fig-intriangle} illustrates the
corresponding geometric construction.

The correctness of this translation follows from the definition of the predicate $\pos$ (see
Definition~\ref{def-triangle-predi-partof}).
    \item The translation of ${\Rtr}_i(\tr{1}, \tr{2}, \ldots, \tr{k})$, where $\trc{i}$ is translated into $\vect{x}_{i,1}$, $\vect{x}_{i,2}$ and $\vect{x}_{i,3}$ for $1 \leq i \leq k$, is
$$\Rpt_i(\vect{x}_{1,1}, \vect{x}_{1,2}, \vect{x}_{1,3}, \vect{x}_{2,1}, \vect{x}_{2,2}, \vect{x}_{2,3}, \ldots, \vect{x}_{k,1}, \vect{x}_{k,2}, \vect{x}_{k,3}).$$

The correctness of this translation follows from Definition~\ref{def-triangle-base} and
Remark~\ref{remark-triangle-can}.
 
\end{enumerate}

\medskip
\par\noindent{\it The translation of composed formulas.}
\medskip
\par\noindent Assume that we already correctly translated the \fom{$\Delta$}{$\sigmatr$}-formulas
$\hat{\varphi}$ and $\hat{\psi}$ into the \fom{$\{\betws\}$}{$\sigmapt$}-formulas $\dot{\varphi}$
and $\dot{\psi}$. Suppose that the number of free
    variables in $\hat{\varphi}$ is $k_{\varphi}$ and that of $\hat{\psi}$ is
    $k_{\psi}$.
Therefor, we can assume that, for each triangle database $\structTr$ over the input schema
$\sigmatr$, and for each $k_{\varphi}$-tuple of triangles $(\trc{1}, \trc{2}, \ldots,
\trc{k_{\varphi}})$ given as $((\vect{a}_{1,1}, \vect{a}_{1,2}, \vect{a}_{1,3}),\ab
(\vect{a}_{2,1}, \vect{a}_{2,2}, \vect{a}_{2,3}), \ab \ldots,\ab (\vect{a}_{k_{\varphi},1},
\vect{a}_{k_{\varphi},2}, \vect{a}_{k_{\varphi},3}))$ that
$$\displaylines{\qquad \structTr\models\hat{\varphi}(\trc{1}, \trc{2}, \ldots, \trc{k_{\varphi}})
\textrm{ if and only if }\splits \structPt\models\dot{\varphi}(\vect{a}_{1,1}, \vect{a}_{1,2},
\vect{a}_{1,3}, \vect{a}_{2,1}, \vect{a}_{2,2}, \vect{a}_{2,3}, \ldots, \vect{a}_{k_{\varphi},1},
\vect{a}_{k_{\varphi},2}, \vect{a}_{k_{\varphi},3})\qquad}$$
 is true when $\structPt$ is the
spatial (point) database  over the input schema $\sigmapt$, obtained from $\structTr$ by applying
the canonical bijection $can_{tr}$ between $\Rnmk{2}{3}{k_{\varphi}}$ and $\Rnm{2}{3k_{\varphi}}$,
on $\structTr$. For the formula $\hat{\psi}$ the analog holds.

In the following, we omit the $k_{\varphi}$-tuples (resp., $k_{\psi}$-tuples) of triangles and
$3k_{\varphi}$-tuples (resp., $3k_{\psi}$-tuples) of points the formulas are applied on, to make
the proofs more readable.

\begin{enumerate}[(i)]
    \item The translation of $\hat{\varphi} \mathrel{\land} \hat{\psi}$ is
    $\dot{\varphi}\mathrel{\land}\dot{\psi}$. Indeed,

    \begin{tabular}{cl}
    & $\structPt \models (\dot{\varphi}\mathrel{\land}\dot{\psi})$ \\
    iff. & $\structPt \models \dot{\varphi}$ and $\structPt \models\dot{\psi}$ \\
    iff. & $\structTr \models \hat{\varphi}$ and $\structTr \models\hat{\psi}$ \\
    iff. & $\structTr \models (\hat{\varphi}\mathrel{\land}\hat{\psi}).$
    \end{tabular}
    \item The translation of $\hat{\varphi} \mathrel{\lor} \hat{\psi}$ is
    $\dot{\varphi}\mathrel{\lor}\dot{\psi}$. Indeed,

    \begin{tabular}{cl}
    & $\structPt \models (\dot{\varphi}\mathrel{\lor}\dot{\psi})$ \\
    iff. & $\structPt \models \dot{\varphi}$ or $\structPt \models\dot{\psi}$ \\
    iff. & $\structTr \models \hat{\varphi}$ or $\structTr \models\hat{\psi}$ \\
    iff. & $\structTr \models (\hat{\varphi}\mathrel{\lor}\hat{\psi}).$
    \end{tabular}
    \item The translation of $\neg\hat{\varphi}$ is
    $\neg\dot{\varphi}$. Indeed,

    \begin{tabular}{cl}
    & $\structPt \models \neg\dot{\varphi}$ \\
    iff. & it is not true that $\structPt \models \dot{\varphi}$\\
    iff. & it is not true that $\structTr \models \hat{\varphi}$\\
    iff. & $\structTr \models \neg\hat{\varphi}.$
    \end{tabular}
    \item Assume that $\hat{\varphi}$ has free variables $\tr{}, \tr{1}, \ldots, \tr{k}$ and $\tr{}$ is translated
    into $\vect{x}_1$, $\vect{x}_2$, $\vect{x}_3$ and $\tr{i}$
    is translated into $\vect{x}_{i,1}$, $\vect{x}_{i,2}$, $\vect{x}_{i,3}$.
    The translation of $$\exists\tr{}\,\hat{\varphi}(\tr{}, \ab\tr{1}, \tr{2}, \ldots, \tr{k})$$
    $$\textrm{is }\exists \vect{x}_1\,\exists \vect{x}_2\,\exists\vect{x}_3\,\dot{\varphi}(\vect{x}_1, \vect{x}_2, \vect{x}_3, \vect{x}_{1,1}, \vect{x}_{1,2}, \vect{x}_{1,3}, \ldots, \vect{x}_{k,1}, \vect{x}_{k,2}, \vect{x}_{k,3}).$$ Indeed,

    \begin{tabular}{cl}
    & $\structPt \models$\\
    &  $\exists\vect{x}_1\,\exists\vect{x}_2\,\exists\vect{x}_3\,\dot{\varphi}(\vect{x}_1, \vect{x}_2, \vect{x}_3)[\vect{a}_{1,1}, \vect{a}_{1,2}, \vect{a}_{1,3}, \ldots, \vect{a}_{k,1}, \vect{a}_{k,2}, \vect{a}_{k,3}]$ \\
    iff. & there exist points $\vect{a}_1$, $\vect{a}_2$, $\vect{a}_3)$ in $\Rn{2}$ such that
    \\
    & $\structPt \models \dot{\varphi}[\vect{a}_1, \vect{a}_2, \vect{a}_3, \vect{a}_{1,1}, \vect{a}_{1,2}, \vect{a}_{1,3}, \ldots, \vect{a}_{k,1}, \vect{a}_{k,2}, \vect{a}_{k,3}]$\\
    iff. & there exists a triangle $\trc{}$ such that

    $\structTr \models \hat{\varphi}[\trc{}, \trc{1}, \ldots, \trc{k}]$, where \\
    & $\trc{i}$ is the triangle with corner points $\vect{a}_{i,1}$, $\vect{a}_{i,2}$ and  $\vect{a}_{i,3}$ for $1 \leq i \leq k$\\
    iff. & $\structTr \models \exists\trc{}\,\hat{\varphi}(\trc{})[ \trc{1}, \ldots, \trc{k}].$
    \end{tabular}
\end{enumerate}
To summarize, let $\sigmatr = \{\Rtr_1, \Rtr_2, \ldots, \Rtr_m\}$ be a spatial triangle database
schema. Let $\sigmapt = \{\Rpt_1, \Rpt_2, \ldots, \Rpt_m\}$ be the corresponding spatial point
database schema. Each formula $\hat{\varphi}$ in \fom{$\Delta$}{$\sigmatr$}, with free variables
$\tr{1}, \tr{2}, \ldots, \tr{k}$ can be translated into a \fom{$\{\betws\}$}{$\sigmapt$}-formula
$\dot{\varphi}$ with free variables $\vect{x}_1, \vect{x}_2, \vect{x}_3, \ab \vect{x}_{1,1},
\vect{x}_{1,2}, \vect{x}_{1,3}, \ab \vect{x}_{2,1}, \vect{x}_{2,2}, \vect{x}_{2,3}, \ab \ldots, \ab
\vect{x}_{k,1}, \vect{x}_{k,2}, \vect{x}_{k,3}$. This translation is such that, for all triangle
databases $\structTr$ over $\sigmatr$, $\structTr \models \hat{\varphi}$ iff. $\structPt \models
\dot{\varphi}$. Here, $\structPt$ is the spatial point database over $\sigmapt$ which is the image
of $\structTr$ under the canonical bijection between $\Rnmk{2}{3}{k}$ and $\Rnm{2}{3k}$. This
completes the soundness proof.\qed
\medskip
For completeness, we translate \foms{$\{\betws\}$}-formulas into \fod-formulas. We again prove this
by induction, on the structure of \foms{$\{\betws\}$}-formulas. This translation  is not as
straightforward as the translation in the other direction, however.

\begin{lemma}[\textbf{Completeness of} \fod]\rm\label{lemma-tr-complete}
Let $\sigmatr = \{\Rtr_1, \Rtr_2, \ldots, \Rtr_m\}$ be a spatial triangle database
    schema and $\sigmapt$ be the corresponding spatial database schema.
    Every \fom{$\{\betws\}$}{$\sigmapt$}-expressible query can be expressed equivalently in \fom{$\Delta$}{$\sigmatr$}.
\end{lemma}
\par\noindent{\bf Proof.}
Let $\sigmatr = \{\Rtr_1, \Rtr_2, \ldots, \Rtr_m\}$ be a spatial triangle database
    schema and $\sigmapt$ be the corresponding spatial database schema. We have to prove that we can translate every triangle database query, expressed in the language
\fom{$\{\betws\}$}{$\sigmapt$}, into a triangle database query in the language
\fom{$\Delta$}{$\sigmatr$} over trangle databases.

We first show how we can simulate point variables by a degenerated triangle, and any
\fom{$\{\betws\}$}{$\sigmapt$}-formula $\dot{\varphi}(\vect{x}_1, \vect{x}_2, \ldots, \vect{x}_k)$
by a formula $\varphi(\tr{1}, \tr{2}, \ldots, \tr{k})$, where $\tr{1}, \tr{2}, \ldots, \tr{k}$
represent triangles that are degenerated into points. We prove this by induction on the structure
of \fom{$\{\betws\}$}{$\sigmapt$}-formulas. Initially, each \fom{$\{\betws\}$}{$\sigmapt$}-formula
$\dot{\varphi}(\vect{x}_1, \vect{x}_2, \ldots, \vect{x}_k)$ will be translated into a
\fom{$\Delta$}{$\sigmatr$}-formula $\hat{\varphi}(\tr{1}, \tr{2}, \ldots, \tr{k})$ with the same
number of free variables.

The translation of a point variable $\vect{x}$ is the triangle variable $\tr{}$, and we add the
condition $\pt{\tr{}}$ as a conjunct to the beginning of the translation of the formula. The
definition of $\pt{\tr{}}$ is
$$\displaylines{\qquad
\forall\tr{}'\,(\po{\tr{}'}{\tr{}}\rightarrow (\tr{} \mathrel{=} \tr{}')). \qquad}$$ In the
following, we always assume that such formulas $\pt{\tr{}}$ are already added to the translation as
a conjunct.
\medskip

\medskip
\par\noindent{\it The translation of atomic formulas.}
\medskip
\par\noindent The atomic formulas of the language \fom{$\{\betws\}$}{$\sigmapt$} are equality constraints on point variables,
 formulas of the form $\betws{(\vect{x}_1, \vect{x}_2, \vect{x}_3)}$, and formulas of the type $\Rpt_i(\vect{x}_1, \vect{x}_2, \ldots,
 \vect{x}_{k})$, where $k =  3\times ar(\Rtr_i)$. We show that all of those can be
 simulated into an equivalent \fom{$\Delta$}{$\sigmatr$} formula.
\begin{figure}
  \centerline{\includegraphics[width=100pt]{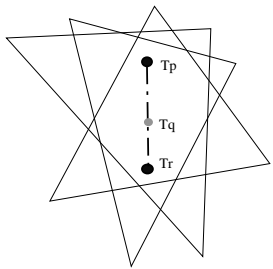}}
  \caption{Illustration of the translation of the predicate $\betws$. The (degenerated) triangle $\trc{q}$ lies between the (degenerated) triangles $\trc{p}$ and $\trc{r}$ if and only if all triangles that contain both $\trc{p}$ and $\trc{r}$, also contain $\trc{q}$.}\label{fig-between-triangles}
\end{figure}
\begin{enumerate}[(i)]
    \item The translation of $(\vect{x}_1 \mathrel{=} \vect{x}_2)$ is $(\tr{1} =_{\vartriangle} \tr{2})$.
    \item The translation of $\betw{\vect{x}_1}{\vect{x}_2}{\vect{x}_3}$, where $\tr{1}$, $\tr{2}$
    and $\tr{3}$ (which as assumed are already declared points) are the translations of $\vect{x}_1$, $\vect{x}_2$ and $\vect{x}_3$, respectively, is expressed by saying that all triangles that contain both $\tr{1}$ and $\tr{3}$
    should also contain $\tr{2}$. It then follows from the convexity of triangles (or line segments, in the degenerated case)
    that $\tr{2}$ lies on the line segment between $\tr{1}$ and $\tr{3}$. Figure~\ref{fig-between-triangles} illustrates this principle. We now give the formula
    translating $\betw{\vect{x}_1}{\vect{x}_2}{\vect{x}_3}$:
    $$\displaylines{\qquad\forall \tr{4}\,((\po{\tr{1}}{\tr{4}}\land \po{\tr{3}}{\tr{4}}) \rightarrow \po{\tr{2}}{\tr{4}}).\qquad}$$

    The correctness of this translation follows from the fact that triangles are convex objects.
    \item  Let $\Rpt_j$ be a relation name from $\sigmapt = \{\Rpt_1, \Rpt_2, \ldots, \Rpt_m\}$. Let $ar(\Rtr_j) = k$ and thus $ar(\Rpt_j) = 3k$, for $1 \leq j \leq m$.
    The translation of $\Rpt_j(\vect{x}_{1,1}, \ab\vect{x}_{1,2}, \ab\vect{x}_{1,3}, \ab\vect{x}_{2,1}, \ab\vect{x}_{2,2}, \ab\vect{x}_{2,3}, \ab\ldots,
\ab\vect{x}_{k,1}, \ab\vect{x}_{k,2}, \ab\vect{x}_{k,3})$ is:
$$\displaylines{\qquad \exists\tr{1}\exists\tr{2}\ldots\exists\tr{k}({\Rtr_j}(\tr{1}, \tr{2}, \ldots, \tr{k}) \mathrel{\land} \bigwedge_{i=1}^k \predi{CornerP}(\tr{i,1}, \tr{i,2}, \tr{i,3}, \tr{i})).\qquad}$$

The definition of $\predi{CornerP}$ is:
$$\displaylines{\qquad\predi{CornerP}(\tr{1}, \tr{2}, \tr{3}, \tr{}):= \forall \tr{4}\,((\pt{\tr{4}} \land \po{\tr{4}}{\tr{}}) \splits \rightarrow \predi{Intriangle}_{\vartriangle}(\tr{4}, \tr{1}, \tr{2}, \tr{3})).\qquad}$$
The predicate $\predi{InTriangle}_{\vartriangle}$ is the translation of the predicate $\predi{InTriangle}$ of the language \foms{$\{\betws\}$}
 as described in the proof of Lemma~\ref{lemma-tr-sound}, into \fod.
The \foms{$\{\betws\}$} formula expressing $\predi{InTriangle}$ only uses $\betws$. In the previous
item of this proof, we already showed how this can be translated into \fod.

Given a $(3k)$-tuple of points $(\vect{a}_{1,1}, \vect{a}_{1,2}, \vect{a}_{1,3}, \ab
\vect{a}_{2,1}, \vect{a}_{2,2}, \vect{a}_{2,3}, \ab \ldots, \ab \vect{a}_{a,1}, \vect{a}_{k,2},
\vect{a}_{k,3})$ in $\Rn{2}$. There will be $(6^k)$ $k$-tuples of triangles  $(\trc{1}, \trc{2},
\ldots, \trc{k})$ such that, for each of the $\trc{i}$, $1 \leq i \leq k$, the condition
$\predi{CornerP}(\trc{i,1}, \trc{i,2}, \trc{i,3}, \trc{i})$ is true. There will, however, only be
one tuple of triangles that is the image of the $(3k)$-tuple of points $(\vect{a}_{1,1},
\vect{a}_{1,2}, \vect{a}_{1,3}, \ab \vect{a}_{2,1}, \vect{a}_{2,2}, \vect{a}_{2,3}, \ab \ldots, \ab
\vect{a}_{a,1}, \vect{a}_{k,2}, \vect{a}_{k,3})$ under the inverse of the canonical bijection
$can_{tr}$. Therefor, the simulation is correct.
\end{enumerate}

\medskip
\par\noindent{\it The translation of composed formulas.}
\medskip

Now suppose that we already simulated the \fom{$\{\betws\}$}{$\sigmapt$} formulas
$\dot{\varphi}(\vect{x}_{1},\ab \vect{x}_{2}, \ldots, \ab\vect{x}_{k_{\varphi}})$ and
$\dot{\psi}(\vect{x}_{1}, \vect{x}_{2}, \ldots, \vect{x}_{k_{\psi}})$ into formulas $\hat{\varphi}$
and $\hat{\psi}$ in \fom{$\Delta$}{$\sigmatr$} with free variables $\tr{{1}}$, $\tr{{2}}$,
$\ldots$, $\tr{{k_\varphi}}$ and $\tr{{1}}'$, $\tr{{2}}'$, $\ldots$, $\tr{{k_\psi}}'$,
respectively. We can hence assume that, for each triangle database $\structTr$ over $\sigmatr$ and
for each $k_{\varphi}$-tuple of triangles ($\trc{{1}}$, $\trc{{2}}$, $\ldots$, $\trc{{k_\varphi}}$)
= $((\vect{a}_{1}, \vect{a}_{1}, \vect{a}_{1}), \ab (\vect{a}_{2}, \vect{a}_{2}, \vect{a}_{2}),
\ab\ldots, \ab(\vect{a}_{k_{\varphi}}, \vect{a}_{k_{\varphi}}, \vect{a}_{k_{\varphi}}))$, which are
required to be degenerated into points, that

$$\displaylines{\qquad\structTr \models \hat{\varphi}[\trc{1}, \trc{2}, \ldots,
\trc{k_{\varphi}}]\textrm{ iff. }\structPt\models \dot{\varphi}[\vect{a}_{1}, \vect{a}_{2}, \ldots,
\vect{a}_{k_{\psi}}]. \qquad}$$ For $\hat{\psi}$ we have analogue conditions.

The composed formulas $\dot{\varphi} \land\dot{\psi}$, $\dot{\varphi} \lor\dot{\psi}$,
$\neg\dot{\varphi}$ and $\exists\vect{x}\,\dot{\varphi}$, are translated into $\hat{\varphi}
\land\hat{\psi}$, $\hat{\varphi} \lor\hat{\psi}$, $\neg\hat{\varphi}$ and $\exists
\tr{}\,(\hat{\varphi})$, respectively if we assume that $\vect{x}$ is translated into $\tr{}$. The
correctness proofs for these translations are similar to the proofs in Lemma~\ref{lemma-tr-sound}.
Therefor, we do not repeat them here. This concludes the proof of Lemma~\ref{lemma-tr-complete}.
\qed

\bigskip

\begin{remark}\label{rem-tr-juiste-ariteit}\rm
So far, we showed that we can simulate any \fom{$\{\betws\}$}{$\sigmapt$} formula
$\dot{\varphi}(\vect{x}_1, \vect{x}_2, \ldots, \vect{x}_k)$ by a formula $\varphi'(\tr{1}, \tr{2},
\ldots, \tr{k})$, where $\pt{\tr{i}}$ is true for all $\tr{i} (1 \leq i \leq k)$. If
$\dot{\varphi}$ expresses a $k$-ary triangle database query $Q$ however (\ie, $\dot{\varphi}$ has
$(3k)$ free variables), we can do better.

Let $\dot{\varphi}$ be the \fom{$\{\betws\}$}{$\sigmapt$}-formula expressing a $k$-ary triangle
database query $\qtr$. The free variables of $\dot{\varphi}$ are $\vect{x}_{1,1}, \vect{x}_{1,2},
\vect{x}_{1,3}, \ab\vect{x}_{2,1}, \vect{x}_{2,2}, \vect{x}_{2,3}, \ab\ldots, \ab\vect{x}_{k,1},
\vect{x}_{k,2}, \vect{x}_{k,3}$.

We now construct the \fom{$\Delta$}{$\sigmatr$} formula $\hat{\varphi}$ expressing the query $\qtr$
as follows:

$$\displaylines{\qquad \hat{\varphi}(\tr{1}, \tr{2}, \ldots, \tr{k}) \equiv\splits\exists\tr{{1,1}}\exists\tr{{1,2}}\exists\tr{{1,3}}\exists\tr{{2,1}}\exists\tr{{2,2}}\exists\tr{{2,3}}\ldots\exists\tr{{k,1}}
\exists\tr{{k,2}}\exists\tr{{k,3}}( \splits \bigwedge_{i=1}^k \predi{CornerP}(\tr{{i,1}},
\tr{{i,2}}, \tr{{i,3}}, \tr{i}) \mathrel{\land}\splits \hat{\varphi}'(\tr{{1,1}}, \tr{{1,2}},
\tr{{1,3}}, \tr{{2,1}}, \tr{{2,2}}, \tr{{2,3}},\ldots, \tr{{k,1}}, \tr{{k,2}}, \tr{{k,3}}) ),
\qquad}$$

For each triple of points, there are $6$ different representations for the triangle having those
points as its corner points. Therefor, for each tuple returned by $\hat{\varphi}'$, $6^k$ tuples
will be returned by $\hat{\varphi}$. But, we know that $\dot{\varphi}$ is a well-defined triangle
query. This means that, for each $(3k)$ tuple of points $((\vect{a}_{1,1},\vect{a}_{1,2},
\vect{a}_{1,3} ), \ab(\vect{a}_{2,1}, \vect{a}_{2,2}, \vect{a}_{2,3}), \ab\ldots,
\ab(\vect{a}_{k,1}, \vect{a}_{k,2}, \vect{a}_{k,3}))$ satisfying $\dot{\varphi}$, also the  tuples
$((\vect{a}_{1,j_{1,1}},\vect{a}_{1,j_{1,2}}, \vect{a}_{1,j_{1,3}} ), \ab(\vect{a}_{2,j_{2,1}},
\ab\vect{a}_{2,j_{2,2}}, \ab\vect{a}_{2,j_{2,3}}), \ab\ldots, \ab(\vect{a}_{k,j_{k,1}},
\ab\vect{a}_{k,j_{k,2}}, \ab\vect{a}_{k,j_{k,3}}))$, where $\sigma_i(1,2,3) =
(j_{i,1},j_{i,2},j_{i,3}) (1 \leq i \leq k; \sigma_i  \in {\cal S}_3)$ and ${\cal S}_3$  is
    the set of all permutations of $\{1,2,3\}$, satisfy $\dot{\varphi}$. Therefor, $\hat{\varphi}$
    and $\dot{\varphi}$ are equivalent according to definition~\ref{def-triangle-query-equiv}.\qed
\end{remark}

\medskip
We now combine the soundness and completeness lemmas, and use them to prove our main theorem for
this section:
\medskip

\begin{theorem}[\textbf{Expressiveness of FO(${\bf \Delta}$)}]\rm\label{theorem-tr-soundcomplete}
Let $\sigmatr = \{\Rtr_1, \Rtr_2, \ldots, \Rtr_m\}$ be a spatial triangle database
    schema. Let $\overline{R}_i$ be the corresponding constraint relation names of
    arity $6\times ar(\Rtr_i)$, for $1 \leq i \leq m$, and let $\overline{\sigma}$ be the spatial database schema $\{\overline{R}_1, \overline{R}_2, \ldots, \overline{R}_m\}$.
The language \fom{${\bf \Delta}$}{$\sigmatr$} is sound and complete for the affine-generic
\fom{\sigPolyNoBr}{$\overline{\sigma}$}-queries on triangle databases.
\end{theorem}
\par\noindent{\bf Proof.}
Let $\sigmatr = \{\Rtr_1, \Rtr_2, \ldots, \Rtr_m\}$ be a spatial triangle database
    schema.
Let $\Rpt_i$ be the corresponding spatial point relation names of
    arity $3\times ar(\Rtr_i)$, for $1 \leq i \leq m$, and let $\sigmapt$ be the spatial database schema $\{\Rpt_1, \Rpt_2, \ldots, \Rpt_m\}$.
    Let $\overline{R}_i (1 \leq i \leq m)$ be the corresponding constraint relation names of
    arity $6\times ar(\Rtr_i)$ and let $\overline{\sigma}$ be the spatial database schema $\{\overline{R}_1, \overline{R}_2, \ldots, \overline{R}_m\}$.

From Lemma~\ref{lemma-tr-sound} and Lemma~\ref{lemma-tr-complete}, we can conclude that
\fom{$\Delta$}{$\sigmatr$} is sound and complete for the \fom{$\{\betws\}$}{$\sigmapt$}-queries on
triangle databases.

Gyssens, Van den Bussche and Van Gucht showed that \fom{$\{\betws\}$}{$\sigmapt$} is sound and
complete for the affine-generic \fom{\sigPolyNoBr}{$\overline{\sigma}$}-queries on geometric
databases~\cite{gvv-jcss}.

From the definition of triangle databases, we know that they are geometric databases. This
concludes the proof. \qed

\medskip
The following remark is important, we will come back to it at the end of this section.
\begin{remark}\label{remark-convexity}\rm 
In the proofs of Lemma~\ref{lemma-tr-sound} and Lemma~\ref{lemma-tr-complete}, we only use the fact
that triangles are convex objects having three corner points. We use no other properties of
triangles.\qed
\end{remark}

\medskip

The following corollary follows from the fact that \fobsigma{\bf +While} is sound and complete for
the computable affine-generic queries on geometric databases~\cite{gvv-jcss}. The language
\fom{${\bf \Delta}$}{$\sigmatr$} {\bf + While} is a language in which \fom{${\bf
\Delta}$}{$\sigmatr$}-definable relations can be created and which has a while-loop with \fom{${\bf
\Delta}$}{$\sigmatr$} -definable stop conditions.

\begin{corollary}[\textbf{Expressiveness of \fom{${\bf \Delta}$}{$\sigmatr$}{\bf + While}}]\rm
Let $\sigmatr$ be a spatial triangle database schema. The language \fom{${\bf
\Delta}$}{$\sigmatr$}{\bf + While} is sound and complete for the computable affine-generic queries
on triangle databases.\qed
\end{corollary}

We now give some examples of \fom{${\bf \Delta}$}{$\sigmatr$}-queries. We illustrate some
geometrical constructions in Example~\ref{ex-tr-geom}. Afterwards, we formulate queries on an
example spatial triangle database in Example~\ref{ex-tr-butterfly}.

\begin{example}\rm\label{ex-tr-geom}
We illustrate how to express that two triangles are similar, \ie, each side of the first triangle
is parallel to a side of the second triangle. We denote the formula expressing this by
$\predi{Sim}$.

We use the predicates $\predi{ColSeg}$ and $\predi{ParSeg}$, expressing that two line segments are
collinear and parallel respectively, to simplify the expression for $\predi{Sim}$.

$$\displaylines{\qquad\predi{ColSeg}(\tr{1}, \tr{2}) := \predi{Seg}(\tr{1}) \land \predi{Seg}(\tr{2}) \land \splits\exists\tr{3}\,(\predi{Seg}(\tr{3}) \land \predi{PartOf}(\tr{1}, \tr{3}) \land \predi{PartOf}(\tr{2}, \tr{3})).\qquad}$$
Here, $\predi{Seg}(\tr{1})$ is a shorthand for
$$\displaylines{\qquad\exists\tr{4}\,\exists\tr{5}\,(\pt{\tr{4}} \land \pt{\tr{5}} \land \splits\forall
\tr{6}\,( (\pt{\tr{6}}  \land  \predi{PartOf}(\tr{6}, \tr{1})) \rightarrow
(\betwtr{\tr{4}}{\tr{6}}{\tr{5}}))).\qquad}$$

The fact that two line segments are parallel is now defined as follows:
$$\displaylines{\qquad\predi{ParSeg}(\tr{1}, \tr{2}):= \predi{Seg}(\tr{1}) \land \predi{Seg}(\tr{2}) \land \forall \tr{3}\,\forall \tr{4}\,(\splits(\predi{ColSeg}(\tr{1}, \tr{3}) \land \predi{ColSeg}(\tr{2}, \tr{4})) \rightarrow \splits \neg\exists\tr{5}\,(\predi{PartOf}(\tr{5}, \tr{3}) \land \predi{PartOf}(\tr{5}, \tr{4}))).\qquad}$$

Now we can write the expression for $\predi{Sim}$:
$$\displaylines{\qquad\predi{Sim}(\tr{1}, \tr{2}) :=\splits
\exists\tr{1,1}\exists\tr{1,2}\exists\tr{1,3}\exists\tr{1,4}\exists\tr{1,5}\exists\tr{1,6}\exists\tr{2,1}
\exists\tr{2,2}\exists\tr{2,3}\exists\tr{2,4}\exists\tr{2,5}\exists\tr{2,6}( \splits\bigwedge_{i =
1}^2(\predi{CornerP}(\tr{i,1}, \tr{i,2}, \tr{i,3}, \tr{i}) \land
\predi{CornerP}(\tr{i,1},\tr{i,1},\tr{i,2},\tr{i,4}) \land \splits
\predi{CornerP}(\tr{i,2},\tr{i,2},\tr{i,3},\tr{i,5}) \land
\predi{CornerP}(\tr{i,3},\tr{i,3},\tr{i,1},\tr{i,6})) \land \splits \bigvee_{\sigma(1,2,3) = (i_1,
i_2, i_3), \sigma \in {\cal S}_3}(\predi{ParSeg}(\tr{1,4}, \tr{2,{(3 + i_1)}}) \land
\predi{ParSeg}(\tr{1,5}, \tr{2,{(3 + i_2)}}) \splits \land \predi{ParSeg}(\tr{1,6}, \tr{2,{(3 +
i_3)}}))),\qquad}$$ where ${\cal S}_3$ is the set of all permutations of $\{1,2,3\}$.\qed 
\end{example}

\medskip
We proceed with an example of a spatial database containing information about butterflies, and some
\fod-queries that can be asked to such a database.

\begin{figure}
\centerline{ \psfig{file=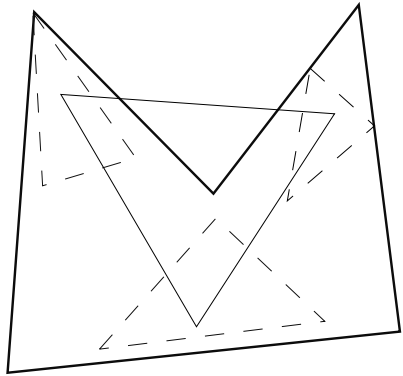,width=150pt}} \caption{The convex hull of a set $S$ of
triangles is computed by adding all triangles constructed from three points that are inside three
triangles of $S$.}\label{fig-convex-tr}
\end{figure}

\begin{example}\rm\label{ex-tr-butterfly}
Consider a triangle database $\structTr$ over the schema $\sigmatr = \{ButterflyB, \ab PlantP,\ab
Rural\}$ that contains information about butterflies and flowers. The unary triangle relation
$ButterflyB$ contains all regions where some butterfly $B$ is spotted. The unary triangle relation
$PlantP$ contains all regions where some specific plant $P$ grows. We also have a unary triangle
relation $Rural$, containing rural regions. It is known in biology that each butterfly appears
close to some specific plant, as caterpillars only eat the leaves of their favorite plant. Suppose
that it is also investigated that butterflies like to live in rural areas.

\medskip
\par\noindent $\bullet$  $Q_{10}:$ {\em Are all butterflies $B$ spotted in regions where the plant $P$ grows?} This
query can be used to see if it is possible that a butterfly was spotted in a certain region. The
query $Q_{10}()$ can be expressed by the formula
$$\displaylines{\qquad \neg(\exists\tr{1}\,\exists\tr{2}\,(ButterflyB(\tr{1})\mathrel{\land}
\predi{RealTriangle}(\tr{2}) \mathrel{\land} \hfill{} \cr  \hfill{} \predi{PartOf}(\tr{2}, \tr{1})
\mathrel{\land} \neg(\exists\tr{3}\,(PlantP(\tr{3}) \mathrel{\land} \predi{PartOf}(\tr{2}, \tr{3})
)))). \qquad}$$

Here, $\predi{RealTriangle}(\tr{})$ is a shorthand for $\neg\predi{Point}(\tr{}) \land
\neg\predi{Line}(\tr{})$.

\medskip
\par\noindent $\bullet$ $Q_{11}:$ {\em Give the region(s) where we have to search if we want to see butterfly
$B$.} The query $Q_{11}(\tr{})$ can be expressed by the formula
 $$\exists\tr{2}\,\exists\tr{3}\,(PlantP(\tr{2}) \mathrel{\land} Rural(\tr{3}) \mathrel{\land}
  \predi{PartOf}(\tr{}, \tr{2}) \mathrel{\land} \predi{PartOf}(\tr{}, \tr{3})).$$

\medskip
\par\noindent $\bullet$ $Q_{12}:$ {\em Give the region inside the convex hull of the search region for butterfly $B$.}
It is much more convenient to search a convex region than having to deal with a very irregularly
shaped region.

We first express how to test whether the region is convex ($Q_{12}'$), this will help understand
the formula that computes the convex hull. The query $Q_{12}'()$ can be expressed by the formula
$$\displaylines{\qquad \forall
\tr{1}\,\forall \tr{2}\,\forall \tr{3}\,\forall \tr{4}\,( (\bigwedge_{i=1}^3 \predi{Point}(\tr{i})
\mathrel{\land} \bigwedge_{i=1}^3 Q_{11}(\tr{i})  \mathrel{\land} \hfill{} \cr  \hfill{}
\predi{CornerP}(\tr{1}, \tr{2}, \tr{3}, \tr{4})) \Rightarrow (Q_{11}(\tr{4}))) . \qquad}$$

The expression
$$\displaylines{\qquad \exists\tr{1}\,\exists\tr{2}\,\exists\tr{3}\,\exists\tr{4}\,\exists\tr{5}\,
\exists\tr{6}\,(\bigwedge_{i=1}^3\predi{Point}(\tr{i}) \mathrel{\land} \hfill{} \cr  \hfill{}
\bigwedge_{i=1}^3\predi{PartOf}(\tr{i}, \tr{i+3}) \mathrel{\land} \bigwedge_{i=4}^6
\predi{Q}_{11}(\tr{i}) \mathrel{\land} \predi{CornerP}(\tr{1}, \tr{2}, \tr{3}, \tr{})) \qquad}$$

hence defines the query $Q_{12}(\tr{})$. For any three points in some triangles in $Q_{11}$, the
triangle connecting them is added to $Q_{12}$. Figure~\ref{fig-convex-tr} illustrates this.\qed
\end{example}

\begin{remark}\rm
The first two queries of Example~\ref{ex-tr-butterfly} ask for relations between regions that can
be expressed by the so-called 9-intersection model~\cite{egherring}. This model defines a relation
between two regions by investigating the intersections between their boundaries, interiors and
exteriors. As the boundary, interior and exterior of a region can be expressed in
\fom{\sigPolyNoBr}{$\overline{\sigma}$}, and are affine invariant concepts~\footnote{To be exact,
they are topological concepts. The affinities of the plane are a subgroup of the homeomorphisms of
the plane, so the invariance under the boundary and interior operations carry over naturally.}, all
relations that can be expressed by the 9-intersection model, can be expressed in \fom{${\bf
\Delta}$}{$\sigmatr$}.\qed
\end{remark}

\begin{remark}\label{remark-generalization}\rm
We now reconsider Remark~\ref{remark-convexity}. In the proofs of Lemma~\ref{lemma-tr-sound} and
Lemma~\ref{lemma-tr-complete}, we only used the fact that triangles are convex objects having three
corner points. It is not difficult to prove that the predicate $\pos$ can be generalized to a
predicate $\posnk{n}{k}$, which arguments are \ndim{n} convex objects with $k$ corner points
(\emph{(n,k)-objects}) and that the language \foms{$\{\posnk{n}{k}\}$} is sound and complete for
the first-order affine-generic queries on $(n,k)$-objects.\qed
\end{remark}

In the context of this remark, we also want to refer to the work of Aiello and van
Benthem~\cite{benthem-2,benthem-1} on modal logics of space. They first propose a topological modal
logic over regions, which can express ``connectedness'' and ``parthood''. By adding a ``convexity''
operator (expressed using a ``betweenness'' operator), they obtain an affine modal logic.
Essentially, we do the same, as triangles are convex and connected sets, and we add the
``parthood'' operator \pos.

In~\cite{benthem-1}, the authors also motivate the use of finite unions of convex sets as basic
elements for spatial reasoning. They argue that it is a very natural way for people to reason about
objects. A fork, for example will be described as the union of its prongs and its handle.

\subsection{Safety of Triangle Database Queries}\label{sec-safety} Triangle
relations can represent infinite sets of triangles. In practice, however, spatial databases will
contain only finite sets of triangles. The $ButterflyB$ and $Rural$ triangle relations of
Example~\ref{ex-tr-butterfly}, for instance, will be modelled in practice using a finite number of
triangles.

The question that arises naturally is whether the language \fod\ returns a finite set of triangles
when the input relations represent finite sets of triangles. The answer is ``no'' (see
Example~\ref{ex-tr-infinite} below). In database theory this problem is usually referred to as the
\emph{safety problem}. Safety of \foms{\sigPolyNoBr}-queries is undecidable in general~\cite{safe},
so we cannot decide {\em a priori} whether a triangle database query will return a finite output or
not.

\medskip
The following example illustrates the fact that the language \fod\ does not necessarily return
finite output on finite input.

\begin{example}\rm\label{ex-tr-infinite}
Let $\sigmatr = \{\Rtr\}$ be a spatial triangle database schema, with $\Rtr$  a triangle relation
containing a finite number of triangles. Consider the following spatial triangle database queries:
\medskip
\par\noindent $\bullet$ $Q_{13}:$ {\em Give all triangles that are part of some triangle of $\Rtr$.}

The query $Q_{13}(\tr{})$ is expressed in \fom{$\Delta$}{$\sigmatr$} by the formula $$\exists
\tr{}'\,(\Rtr(\tr{}') \mathrel{\land} \predi{PartOf}(\tr{}, \tr{}')).$$
\par\noindent $\bullet$ $Q_{14}:$ {\em Give all triangles that intersect some triangle of $\Rtr$.}
The query $Q_{14}(\tr{})$ can be expressed by the formula
$$\exists\tr{}'\,(\Rtr(\tr{}') \mathrel{\land}\predi{Intersect}(\tr{}, \tr{}')).$$

\par\noindent $\bullet$ $Q_{15}:$ {\em Give all the corner points of all triangles of $\Rtr$.}
The query  $Q_{15}(\tr{})$ can be expressed by the formula
$$\displaylines{\qquad \exists\tr{1}\,\exists\tr{1,2}\,\exists\tr{1,3}\,(\Rtr(\tr{1}) \mathrel{\land} (\predi{CornerP}(\tr{}, \tr{1,2},
\tr{1,3}, \tr{1}) \hfill{} \cr \hfill{} \mathrel{\lor} \predi{CornerP}(\tr{1,2}, \tr{}, \tr{1,3},
\tr{1}) \mathrel{\lor} \predi{CornerP}(\tr{1,2}, \tr{1,3}, \tr{}, \tr{1})) ). \qquad}$$
\par\noindent The queries $Q_{13}$ and $Q_{14}$  return an infinite set of triangles. The query $Q_{15}$ returns a finite number of triangles on the condition that the input relation $\Rtr$ is finite. \qed
\end{example}

As we cannot decide whether a given triangle database query will return a finite result, we turn to
the question of determining whether the result of the query is finite or not, after executing the
query. The answer is affirmative:

\begin{proposition}[\textbf{Finiteness of triangle relations is decidable}]\rm\label{prop-tr-decidable-finite}
It is decidable whether a triangle relation consists of a finite number of triangles. Moreover,
there exists a \foms{$\Delta$}{$\{\Rtr\}$} query that decides whether the triangle relation named
$\Rtr$ consists of a finite number of triangles.
\end{proposition}
\par\noindent{\bf Proof.}
A triangle relation of arity $k$ corresponds to a \sa set in $\Rn{6k}$. The canonical bijection
$can\circ can_{tr}: \Rnmk{2}{3}{k} \rightarrow \Rn{6k}$ establishes this correspondence. A triangle
relation is finite if and only if the corresponding \sa set contains a finite number of points (in
$\Rn{(6k)}$). It is well known that there exists a \foms{\sigPolyNoBr}-formula deciding whether a
\sa set contains a finite number of points. Also, the fact that a triangle relation contains a
finite number of $k$-tuples of triangle is affine-invariant. From the fact that the property is
affine-invariant and expressible in \foms{\sigPolyNoBr}, it follows (from
Theorem~\ref{theorem-tr-soundcomplete}) that there is a a \fom{$\Delta$}{$\{\Rtr\}$}-formula
expressing whether a triangle relation $\Rtr$ is finite or not.\qed

\medskip

We now have a means of deciding whether a triangle relation is finite, but it seems this
requirement is too restrictive.

In Definition~\ref{def-tr-drawing} in Section~\ref{sec-definitions}, we introduced the concept
\emph{drawing} of a triangle. We now straightforwardly extend this definition to spatial triangle
databases.

 \begin{definition}[\textbf{Drawing of a triangle relation}]\label{def-tr-drawing-rel}\rm
Let $\Rtr$ be a triangle relation of arity one. The {\em drawing} of $\Rtr$ is the two-dimensional
figure that is the union of the drawings of all triangles in $\Rtr$.\qed
\end{definition}

For the remainder of this text, we restrict triangle relations (and triangle database queries) to
be unary. It is not clear immediately if it would make sense to define drawings on relations or
queries with an arity greater than one. For example, consider a binary relation containing only one
tuple of line-adjacent non-degenerated triangles. If we draw this relation, we would like to draw
both triangles participating in the relation. This gives the same result as the drawing of a unary
relation containing two tuples. So the drawing apparently ``wipes out'' the relationship between
the triangles.

We also remark the following.

\begin{remark}\label{remark-tr-same-drawing}\rm
Different triangle relations can have the same drawing. Therefore, it seems natural to extend the
strict notion of finiteness of a triangle relation to the existence of a finite triangle relation
having the same drawing. Query $Q_1$ from Example~\ref{ex-tr-infinite}, for instance, seems to be a
query we would like to call ``finite'', because there exists a finite union of triangles with the
same drawing. Indeed, the drawing of the union of all triangles that are part of a given triangle,
is the same as the drawing of the given triangle itself. Query $Q_2$ clearly returns an infinite
set of triangles that is cannot be represented as a finite union of triangles. This is the type of
query we don't want to allow.\qed
\end{remark}

Fortunately, given the output of a unary query, we can determine whether its drawing can be
represented as a finite union of triangles.

\begin{proposition}[\textbf{Finite triangle representation}]\rm\label{prop-tr-finite-rep}
Let $\sigmatr$ be a spatial triangle database schema. Given a unary triangle database query $\qtr$
that is expressible in \fom{$\{\pos\}$}{$\sigmatr$} and a spatial triangle database $\structTr$
over $\sigmatr$, it is decidable whether the unary relation, named $\Rtr_{\qtr}$, containing
$\qtr(\structTr)$ can be represented as a finite union of triangles. Furthermore, there exists a
\fom{$\Delta$}{$\sigmatr'$}-formula deciding this for $\sigmatr' = \sigmatr \cup \{\Rtr_{\qtr}\}$.
\end{proposition}
\par\noindent{\bf Proof.}
It is clear that if the drawing of a triangle relation can be represented as a finite union of
triangles, it can be represented by a \foms{\sigPolyNoBr}-formula using only polynomials of degree
at most one. A set that can be described using polynomials of at most degree one, is called a
semi-linear set. It is well-known that the bounded semi-linear sets are the same as finite unions
of bounded polytopes (which triangles are).

So, if we can check whether the drawing of a (possibly infinite) set of triangles is bounded and
can be represented using polynomials of degree at most one, we know that the set can be represented
by a finite number of triangles.

Checking whether the drawing of a triangle relation $\Rtr$ is bounded can be done easily in
\foms{$\{\pos\}$}{$\{\Rtr\}$}. The following formula performs this check.
$$\displaylines{\qquad\predi{IsBounded}() := \exists\tr{1}\,\forall\tr{2}\,( \Rtr(\tr{2})
\rightarrow \predi{PartOf}(\tr{2}, \tr{1})) .\qquad}$$

Also, we can decide whether a two-dimensional\footnote{Note that this is not true for arbitrary
dimensions.} \sa set can be represented using polynomials of degree at most $k$, for any natural
number $k$~\cite{ks_cdb97}. There exists a \foms{\sigPolyNoBr}-formula deciding
this~\cite{ks_cdb97}. It is clear that the drawing of a unary triangle relation is a \sa set.

From the facts that $(i)$ computing the drawing of a triangle relation is an affine-generic query
that can be expressed in \foms{\sigPolyNoBr} and that $(ii)$ checking whether a triangle relation
has a bounded drawing can be expressed in \fom{$\{\pos\}$}{$\sigmatr$} and that $(iii)$ there
exists a \foms{\sigPolyNoBr}-formula deciding whether the drawing of a triangle relation can be
expressed by polynomials of degree at most one can be done in \foms{\sigPolyNoBr} and, finally,
that $(iv)$ the fact that the drawing of a triangle relation can be expressed by polynomials of
degree at most one is affine-invariant, we conclude that we can decide whether a triangle relation
has a finite representation, and that we can construct a \fom{$\{\pos\}$}{$\sigmatr$}-formula
deciding this.\qed

\medskip
\par\noindent
We now show that, if the drawing of the output of a triangle database query is representable as a
finite set of triangles, we can compute such a finite triangle representation in
\fom{$\{\pos\}$}{$\sigmatr$}.

In \cite{HK-ACMGIS-04}, we proposed an algorithm that computes an affine invariant triangulation of
a set of triangles. Recall that this algorithm computes the drawing of the input triangles, then
partitions this drawing into a set of convex polygons according to the carriers of its boundary
segments and finally triangulates convex polygons by connecting their center of mass to their
corner points.

We assumed in \cite{HK-ACMGIS-04} that the input set of triangles for the triangulation algorithm
was finite. On an infinite collection of triangles for which there exists a finite collection of
triangles with the same drawing, this algorithm would work also correctly, however. The
triangulation described in \cite{HK-ACMGIS-04} therefor seems a good candidate for representing
infinite sets of triangles by finite sets of triangles. But, in~\cite{HK-ACMGIS-04}, we conjectured
that this triangulation cannot be expressed in \fom{$\{\pos\}$}{$\{\Rtr\}$}. The reason for this is
the conjecture 
that the center of mass of a polygon, which is an
affine-invariant, cannot be expressed in \foms{\sigPolyNoBr}, and therefore, also not in
\fom{$\{\pos\}$}{$\{\Rtr\}$}.

\begin{conjecture} \rm\label{conj-centerom} 
Let $P = \{\vect{a}_1, \vect{a}_2, \ldots, \vect{a}_k\}$ be a set of corner points that represent a
convex polygon. Assume that $k
> 3$. The center of mass of the polygon represented by $P$ cannot be expressed in \foms{\sigPolyNoBr}.\qed
\end{conjecture}

Remark that the center of mass of an arbitrary set of points is not expressible in
\foms{\sigPolyNoBr}.

So, the triangulation algorithm from \cite{HK-ACMGIS-04} cannot be used. But, this algorithm
computes a \emph{partition}\xspace of the input into triangles, which is not a requirement here. If
we relax the requirement of having a partition of the original figure into triangles down to having
a {\em finite union} of (possibly overlapping) triangles representing the figure, we can avoid the
computation of the center of mass. The adapted algorithm 
$AfTr(S)$ in given in Figure~\ref{algo-tr-info}.

\begin{figure} 
\begin{algorithmic}[1]
 \REQUIRE{$S$ is a unary triangle relation that can be represented as a finite union of triangles.}
 \STATE Compute the boundary $B_S$ of $S$. $B_S$ is a finite set
of line segments and points.
 \STATE Compute the set of carriers for all line segments of $B_S$. Those carriers partition $S$ into
 a finite union of open convex polygons, points and open line segments. All closures of line segments that do not form a
 side of one of the convex polygons, together with all points that are not a corner point of one of
 the convex polygons are returned as degenerated triangles. Remark that we can return the closures
 of the line segments as $S$ originally is a union of closed triangles, closed line segments and
 points.
\FOR {each polygon}

\STATE output the finite set of triangles that connect three distinct corner points of the polygon
\ENDFOR
\end{algorithmic}\caption{The algorithm $AfTr(S)$.}\label{algo-tr-info}
\end{figure}

Given an unary triangle relation $\Rtr$, we denote the result of algorithm $AfTr(S)$ in Figure~\ref{algo-tr-info} on
input $\Rtr$ by the {\em affine finite triangle representation} of $\Rtr$, or, abbreviated,
$AfTr(\Rtr)$. Now we show that $AfTr(\Rtr)$ can be computed in \fom{${\bf Delta}$}{$\Rtr$},
provided that $\Rtr$ can be represented as a finite union of triangles.

\begin{proposition}[\textbf{Affine finite triangle representation}]\rm
Given a unary triangle relation $\Rtr$ that can be represented as a finite union of triangles, then
there exists an \fom{${\bf \Delta}$}{$\{\Rtr\}$}-formula returning $AfTr(\Rtr)$.
\end{proposition}
\par\noindent{\bf Proof.} We use the fact that all affine-generic semi-algebraic queries on triangle databases can be
expressed in \fom{${\bf Delta}$}{$\Rtr$}.  Therefor, we have to prove that, first, the affine
finite triangle representation is affine-invariant and, second, that the affine finite
representation is expressible in \foms{\sigPolyNoBr}.
\medskip
\par\noindent{\it The affine finite representation is an affine invariant.}
\medskip
\par\noindent We only have to prove this for Step 3 of algorithm $AfTr(S)$ in Figure~\ref{algo-tr-info}. The rest
follows from the analogous property in  \cite{HK-ACMGIS-04}.

Let $\{\vect{a}_1, \vect{a}_2, \ldots, \vect{a}_k\}$ be the set of corner points of a convex
polygon $P$, where $k \geq 3$. Let $\alpha$ be an affinity of the plane. The set
$\{\alpha(\vect{a}_1), \alpha(\vect{a}_2), \ldots, \alpha(\vect{a}_k)\}$ contains the corner points
of the convex polygon $\alpha(P)$. It is clear that, for each triangle $(\vect{a}_h, \vect{a}_i,
\vect{a}_j)$ (such that $h\neq i$, $i\neq j$, $h\neq j$ and $1 \leq h, i, j \leq k$) connecting
three corner points of $P$, the triangle $\alpha(\vect{a}_h, \vect{a}_i, \vect{a}_j) =
(\alpha(\vect{a}_h), \alpha(\vect{a}_i), \alpha(\vect{a}_j))$ is an element of the set of triangles
connecting three corner points of $\alpha(P)$.
\medskip
\par\noindent{\it The affine finite representation is computable in} \foms{\sigPolyNoBr}.
\medskip
\par\noindent In \foms{\sigPolyNoBr}, it is possible to compute the boundary of a semi-linear set (Line 1 of the algorithm $AfTr(S)$ in Figure~\ref{algo-tr-info}). It is also
possible to compute the carriers of all boundary line segments, and their intersection points (Line
2). It can be expressed that two points belong to the same convex polygon, namely, by expressing
that the line segment in between them is not intersected by a carrier. Finally, the set of all
triples of intersection points between carriers that belong to the same convex polygon can be
computed in \foms{\sigPolyNoBr} (Lines 3 through 5). From the fact that the triangle representation
is affine invariant and computable in \foms{\sigPolyNoBr}, it follows that it is computable in
\fod.\qed

\medskip

\medskip
This section on safety finishes the ``spatial'' part of this text. In the remaining part, we
develop a query language for \st triangle databases.

\section{Spatio-temporal Triangle Queries}\label{sec-tr-logic-st}

In this section, we will extend the spatial triangle logic \foms{$\{\pos\}$} to a logic over
spatio-temporal triangles, \ie, triples of \cotemp points in  $\RnR{2}$. The genericity classes we
consider in this section, are the group $({\cal A}_{st}, {\cal A}_t)$ of time-dependent affinities,
the group $({\cal V}_{st}, {\cal A}_t)$ of velocity-preserving transformations and the group
$({\cal AC}_{st}, {\cal A}_t)$ of acceleration-preserving transformations. The first group is a
natural \st extension of the affinities of space. We also include the two other groups, because
they are very relevant from a practical point of view, and because the point languages we
previously identified as generic for those groups were not very intuitive.

Recall that ${\cal A}_t$  is the group of the affinities on the time line and that the elements of
${\cal A}_{st}$ are of the form

$$\left( \begin{array}{c} x_1\\  x_2 \\ \vdots\\  x_n\\ t
 \end{array}
 \right)\mapsto
\left(
\begin{array}{cccc }
\alpha_{11}(t) & \alpha_{12}(t) & \cdots & \alpha_{1n}(t)   \\
\alpha_{21}(t) & \alpha_{22}(t) & \cdots & \alpha_{2n}(t)   \\
\vdots&\vdots&\cdots& \vdots
\\ \alpha_{n1}(t) & \alpha_{n2}(t) & \cdots & \alpha_{nn}(t)
 \end{array}
 \right)\cdot  \left(
\begin{array}{c}
x_1\\  x_2 \\ \vdots\\  x_n
 \end{array}
 \right) + \left(
\begin{array}{c}
\beta_1(t)\\  \beta_2(t)\\\vdots\\ \beta_n(t)\\
 \end{array}
 \right),$$

 where the matrix of the $\alpha_{ij}(t)$ is an affinity for each value of $t$. The group $({\cal AC}_{st}, {\cal
 A}_t)$ is the subgroup of $({\cal A}_{st}, {\cal A}_t)$ in which the functions $\alpha_{ij}$ are
 constants and the functions $\beta_{ij}$ are linear functions of time. The group $({\cal V}_{st}, {\cal
 A}_t)$ is the subgroup of $({\cal AC}_{st}, {\cal A}_t)$ where the $\beta_{ij}$ are constants too.

In \cite{ghk-01}, we proposed point languages capturing exactly those genericity classes.
Table~\ref{table-repeat-point-lang} summarizes the point languages expressing all $({\cal F}_{st},
{\cal F}_t)$-generic queries, for the above groups $({\cal F}_{st}, {\cal F}_t)$. As we will always
assume, in this section, that the underlying dimension is $2$, we adapted the table accordingly.
Now we propose \st point languages that have the same expressivity as the languages listed in
Table~\ref{table-repeat-point-lang}, but on \st triangle databases.

\begin{table}[htbp]
  \begin{center}
    \leavevmode
    \begin{tabular}[c]{|c|l|}
\hline
 \ \      $({\cal F}_{st}, {\cal T}_t)$ \ \ & \ \   Set of point predicates $\Pi{({\cal F}_{st}, {\cal T}_t)}$\\
\hline\hline \ \ $({\cal A}_{st}, {\cal A}_t)$\ \ & \ \ $\{\betwsCotemp,\befores, \eqcrsts \}$\\  \
\ $({\cal
AC}_{st}, {\cal A}_t)$\ \ & \ \ $\{\betws,\befores \}$\\
\ \ $({\cal V}_{st}, {\cal A}_t)$\ \  & \ \
$\{\betws,\befores, \predi{EqSpace} \}$\\
 \hline
    \end{tabular}
\medskip
    \caption{The point logics \foms{$\Pi{({\cal F}_{st}, {\cal T}_t)}$} capturing the \fo $({\cal F}_{st}, {\cal T}_t)$-generic queries, for the classes $({\cal A}_{st}, {\cal A}_t)$, $({\cal AC}_{st}, {\cal A}_t)$ and $({\cal V}_{st}, {\cal A}_t)$. }
    \label{table-repeat-point-lang}
  \end{center}
\end{table}

We will start with the most general transformation group, the group $({\cal A}_{st}, {\cal A}_t)$
of time-dependent affinities.

\subsection{Predicates Invariant under Time-dependent Affinities}\label{sec-st-predi}
In this section, we propose a set of \st triangle predicates such that the spatio-temporal triangle
logic with this predicate set, captures exactly the $({\cal A}_{st}, {\cal A}_t)$-generic queries
on \st triangle databases that are expressible in \foms{\sigPolyNoBr}. We can prove this by
comparing the expressiveness of this \st triangle logic with the language \fobST, when used as a
\st triangle query language (see Definition~\ref{def-triangle-fob-is-poss-lang}). Recall also that
we will have to make sure that the result of a \st triangle query is a consistent \st triangle
relation.

\medskip
The nature of the class $({\cal A}_{st}, {\cal A}_t)$ is such that $({\cal A}_{st}, {\cal
A}_t)$-generic queries can describe snapshots of a \st database in fairly much detail, \ie, all
affine-invariant properties of the snapshot can be expressed. In between snapshots, the expressive
power of $({\cal A}_{st}, {\cal A}_t)$-generic queries is more limited. This follows directly from
the fact that an element of $({\cal A}_{st}, {\cal A}_t)$ transforms each snapshot with another
affinity. We now want to construct a $({\cal A}_{st}, {\cal A}_t)$-generic query language for \st
triangle databases. This means we will be able to describe a \st triangle database by means of its
snapshots, which are collections of snapshots of \st triangles in $(\Rn{2} \times\{\tau_0\})^3$,
for some $\tau_0 \in \R$. The  basic objects for our new language will be, accordingly, triples of
\cotemp points. In this section, we will call these triples of points \emph{triangle snapshots}.
Triangle snapshot variables will be denoted $\trST{}, \trST{1}, \trST{2}, \ldots$ and triangle
snapshot constants by $\trcST{}, \trcST{1}, \trcST{2}, \ldots$. If we want to emphasize the
connection between a triangle snapshot and its corner points, we use the notation $\trcST{pqr}$.

In our search for a set of predicates on triangle snapshots for a $({\cal A}_{st}, {\cal
A}_t)$-generic query language, or, a language with the same expressive power as the language \fobST
on \st triangle databases, the following observations are helpful.

\begin{enumerate}[(i)]
    \item In \cite{ghk-01}, we showed that we need the
binary predicate $\befores$ on points to reflect the monotonicity of time, which is preserved by
the transformation group $({\cal A}_{st}, {\cal A}_t)$.
    \item The predicate $\betwsCotemp$ is used to express affine-invariant properties of
    \cotemp points.
    \item In Section~\ref{sec-tr-logic-spatial}, we showed that
    the predicate \pos\xspace has the same expressive power as the
    predicate $\betws$, on (spatial) triangles.
\end{enumerate}

From observation (i) it follows that the query language we want to construct should be able to
express the order on triangle snapshots. We introduce the triangle snapshot predicate $\beforetrs$,
which, when applied to two triangle snapshots, expresses that the first one is strictly before or
co-temporal with the second one. We will define this more formally later.

From observation (ii) and (iii), we conclude that we can use, slightly adapted, the predicate \pos
on \cotemp triangle snapshots, we will denote it $\posCotemp$. This will allow us to express
snapshots of \st triangle databases in an affine-invariant way. Concluding, the set of \st triangle
predicates we are looking for should contain the elements $\posCotemp$ and $\beforetrs$. Because,
in the end, we want to express al queries expressible in \fobST, on \st triangle databases, we
still have to look for a (set of) triangle snapshot predicate(s) capturing the expressive power of
the predicate $\eqcrsts$.

We repeat the definition of the point predicate $\eqcrsts$. For six \st points $p_1, p_2, p_3,\ab
q_1, q_2, q_3 \in \ab \RnR{2}$, $\eqcrst{p_1}{p_2}{p_3}{q_1}{q_2}{q_3}$ expresses that the
cross-ratio of the three \cotemp and collinear points $p_1$, $p_2$ and $p_3$ equals the cross-ratio
of the time coordinates $\tau_{q_1}$, $\tau_{q_2}$ and $\tau_{q_3}$ of the points $q_1$, $q_2$ and
$q_3$. The expression $\eqcrst{p_1}{p_2}{p_3}{q_1}{q_2}{q_3}$ implicitly refers to a movement.
Indeed, the line segment defined by the points $p_1$ and $p_3$ and the interval $[\tau_{q_1},
\tau_{q_3}]$ can be interpreted as the spatial and temporal projection of a linear movement with
constant speed and we can then interpret $\eqcrst{p_1}{p_2}{p_3}{q_1}{q_2}{q_3}$ as an expression
of the fact that when an object moves with constant speed from $p_1$ to $p_3$ during the interval
$[\tau_{q_1}, \tau_{q_3}]$, it passes
$p_2$ at time moment $\tau_{q_2}$.  

\medskip

There is one obvious way to define the speed of a moving point. For moving triangles, or moving
objects in general, the definition of \emph{speed} is somewhat ambiguous. Triangles can move by
changing their position, but also by changing their shape. We define the speed (resp.,
acceleration) of a moving triangle as the speed (resp., acceleration) of is moving \emph{center of
mass}. Hence, a triangle that is growing or shrinking, but its center of mass remains in the same
position, has zero speed. Based on that definition, we propose a \st triangle database query
language, with the triangle predicates $\posCotemp$, $\beforetrs$ and $\predi{Cas}$ (which is an
abbreviation of ``Constant Average Speed''). The predicate $\predi{Cas}$ takes six arguments
$\trST{1}, \trST{2}, \ldots, \trST{6}$. The first three triangle snapshots, $\trST{1}, \trST{2}$
and $\trST{3}$, are \cotemp and their barycenters are collinear. The last three triangle snapshots,
$\trST{4}, \trST{5}$ and $\trST{6}$, indicate three different time moments. Furthermore, the
cross-ratio of the barycenters of $\trST{1}, \trST{2}$ and $\trST{3}$ is the same as the
cross-ratio of the time coordinates of $\trST{4}, \trST{5}$ and $\trST{6}$. Intuitively, this
predicate, similar to the point predicate $\eqcrsts$, approximates or estimates a linear movement.
Given the time interval during which a triangle moves from the first position to the second one, it
estimates, assuming the triangle moves with constant speed, how long it will take to reach the
position of the third triangle.

It turns out, however, that the language with these three triangle predicates is not very intuitive
to express properties of the shape of triangles, e.g., their relative areas. Therefor, we will also
propose an alternative language. This language has exactly the same expressivity as the first one,
but offers a more direct means to express shape properties of triangles. We propose to replace the
predicate $\predi{Cas}$ by the predicate $\predi{Lex}$ (which is an abbreviation for ``Linear
Expansion''). This predicate also takes six arguments $\trST{1}$, $\trST{2}$, $\ldots$, $\trST{6}$.
The first three triangle snapshots, $\trST{1}, \trST{2}$ and $\trST{3}$, are \cotemp and both
$\poCotemp{\trST{1}}{\trST{2}}$ and $\poCotemp{\trST{2}}{\trST{3}}$ hold. The other three triangle
snapshots exist at three different time moments. Finally, the cross-ratio of the time coordinates
of $\trST{4}, \trST{5}$ and $\trST{6}$ equals the cross ratio of the areas of the three first
triangles. More exactly, $$\frac{|A(\trST{2}) - A(\trST{1})|}{|A(\trST{3}) - A(\trST{1})|} =
\frac{|\tau_{\trST{2}} - \tau_{\trST{1}}|}{|\tau_{\trST{3}} - \tau_{\trST{1}}|},$$ where
$\tau_{\trST{i}}$ denotes the time moment at which $\trST{i}$ exists and $A(\trST{i})$ denotes the
area of the triangle $\trST{i}$. Intuitively, this predicate approximates or estimates a linear
growth or expansion. Given the time interval during which the first triangle expanded into the
second one, it estimates, assuming the triangle grows linearly, how long it will take to reach the
area of the third triangle.

In applications where objects are not growing or shrinking, a language with the predicate
$\predi{Cas}$ may be preferred, whereas in applications where objects do change their shape, the
predicate $\predi{Lex}$ may be preferred. Of course, one can also include both predicates to make
the language suitable for all types of applications.

We will prove that the both the languages \foms{$\{\posCotemp,\ab \beforetrs, \ab\predi{Cas} \}$}
and \foms{$\{\posCotemp, \ab\beforetrs, \ab\predi{Lex} \}$} are sound and complete for the $({\cal
A}_{st}, {\cal A}_t)$-generic first-order \st database queries.

\subsubsection{Expressiveness of the Language \foms{$\{\posCotemp,\ab \beforetrs, \ab\predi{Cas}
\}$}}

In this section, we first give the definitions of the triangle predicates $\posCotemp$,
$\beforetrs$ and $\predi{Cas}$. Next, we show that the language \foms{$\{\posCotemp,\ab \beforetrs,
\ab\predi{Cas} \}$} produces queries that are well-defined on \st triangle databases. After that,
we show its expressive power.  

\begin{definition}[\textbf{The triangle snapshot predicate $\posCotemp$}]\label{def-triangle-st-predi-partof}\rm
Let $\trcST{1} = (p_{1,1}, \ab p_{1,2}, \ab p_{1,3})$ and $\trcST{2} = (p_{2,1}, p_{2,2}, p_{2,3})$
be two triangle snapshots. The binary predicate $\posCotemp$, applied to $\trcST{1}$ and
$\trcST{2}$ expresses that $p_{1,1}$, $p_{1,2}$ and $p_{1,3}$ (resp., $p_{2,1}$, $p_{2,2}$ and
$p_{2,3}$) are \cotemp and that the convex closure of the three points $p_{1,1}$, $p_{1,2}$ and
$p_{1,3}$ is a subset of the convex closure of the three points $p_{2,1}$, $p_{2,2}$ and $p_{2,3}$.\qed
\end{definition}

\begin{definition}[\textbf{The triangle snapshot predicate $\beforetrs$}]\label{def-triangle-st-predi-before}\rm
Let $\trcST{1}\ab  = (p_{1,1}, \ab p_{1,2}, \ab p_{1,3})$ and $\trcST{2} = (p_{2,1}, p_{2,2}, p_{2,3})$
be two triangle snapshots. The binary predicate $\beforetrs$, applied to $\trcST{1}$ and
$\trcST{2}$ expresses that $p_{1,1}$, $p_{1,2}$ and $p_{1,3}$ (resp., $p_{2,1}$, $p_{2,2}$ and
$p_{2,3}$) are \cotemp and that the time coordinate $\tau_{p_{1,1}}$ of $p_{1,1}$ is smaller than
or equal to the time coordinate $\tau_{p_{2,1}}$ of $p_{2,1}$.\qed
\end{definition}

\begin{definition}[\textbf{The triangle snapshot predicate $\predi{Cas}$}]\label{def-triangle-st-predi-cas}\rm
 Let $\trcST{1} = (p_{1,1}, \ab p_{1,2}, \ab p_{1,3}), \ab\trcST{2} =
(p_{2,1}, p_{2,2}, p_{2,3}), \ab\ldots, \ab\trcST{6} = (p_{6,1}, p_{6,2}, p_{6,3})$ be six triangle
snapshots. Let $q_{1}$ (resp., $q_2$, $q_3$) be the barycenter of $\trcST{1}$ (resp., $\trcST{2}$,
$\trcST{3}$). The $6$-ary predicate $\predi{Cas}$, applied to $\trcST{1}$, $\trcST{2}$, \ldots,
$\trcST{6}$ expresses that $p_{i,1}$, $p_{i,2}$ and $p_{i,3}$ are \cotemp for $i = 1 \ldots 6$,
that $q_1$, $q_2$ and $q_3$ are collinear and that the cross-ratio of the points $q_1$, $q_2$ and
$q_3$ is the same as the cross-ratio of the time coordinates $\tau_{p_{4,1}}$, $\tau_{p_{5,1}}$ and
$\tau_{p_{6,1}}$ of $p_{4, 1}$, $p_{5,1}$ and $p_{6,1}$, respectively.\qed
\end{definition}

We now show, by induction on their structure, that the \foms{$\{\posCotemp,\ab \beforetrs,
\ab\predi{Cas} \}$}-queries are well-defined on \st triangle databases.

\begin{lemma}[\textbf{\foms{$\{\posCotemp,\ab \beforetrs, \ab\predi{Cas} \}$} is well-defined}]\rm\label{lemma-tr-st-cas-well-defined}
Let $\sigmatrST = \{\RtrST_1, \ab\RtrST_2, \ab\ldots, \ab\RtrST_m\}$ be a \st triangle database
    schema. Let $\structTrST$ be a consistent \st triangle database over $\sigmatrST$.
For each \fom{$\{\posCotemp,\ab \beforetrs, \ab\predi{Cas} \}$}{$\sigmatrST$}-query $\qtr$,
$\qtr(\structTrST)$ is a consistent triangle relation.
\end{lemma}
\par\noindent{\bf Proof.} Let $\sigmatrST = \{\RtrST_1, \ab\RtrST_2, \ab\ldots, \ab\RtrST_m\}$ be a \st triangle database
    schema. Let $\structTrST$ be a consistent spatial triangle database over $\sigmatrST$.

We prove this lemma by induction on the structure of \fom{$\{\posCotemp,\ab \beforetrs,
\ab\predi{Cas} \}$}{$\sigmatr$}-queries. The atomic formulas of \fom{$\{\posCotemp,\ab \beforetrs,
\ab\predi{Cas} \}$}{$\sigmatr$} are equality expressions on \st triangle variables, expressions of
the form $\poCotemp{\trST{1}}{\ab\trST{2}}$, expressions of the form
$\beforetr{\trST{1}}{\ab\trST{2}}$, expressions of the form $\predi{Cas}(\trST{1},\ab\trST{2},
\ab\ldots, \ab\trST{6})$, and expressions of the form ${\RtrST_i}(\trST{1}, \trST{2}, \ldots,\ab
\trST{ar(\Rtr_i)})$, where $\RtrST_i (1 \leq i \leq m)$ is a relation name from $\sigmatrST$. More
complex formulas can be constructed using the Boolean operators $\land$, $\lor$ and $\neg$ and
existential quantification.

For the atomic formulas, it is easy to see that, if two triangles $\trcST{1}$ and $\trcST{2}$
satisfy the conditions $\trcST{1} =_{\vartriangle} \trcST{2}$, $\poCotemp{\trcST{1}}{\trcST{2}}$,
or $\beforetr{\trcST{1}}{\trcST{2}}$ that also $\trcST{3} =_{\vartriangle} \trcST{4}$ respectively
$\poCotemp{\trcST{3}}{\trcST{4}}$, $\beforetr{\trcST{3}}{\trcST{4}}$ are true if and only if $\trc{1}
=_{\vartriangle}\trc{3}$ and $\trc{2}=_{\vartriangle}\trc{4}$ are true. As we assume the input
database $\structTr$ to be consistent, the atomic formulas of the type $\RtrST_i(\trST{1},
\trST{2}, \ldots, \trST{ar(\RtrST_i)})$, where $(1 \leq i \leq m)$, trivially return consistent
triangle relations.

For the predicate $\predi{Cas}$, the proof is less straightforward. First, it is true that any pair
of triangles $\trcST{}$ and ${\trcST{}}'$ such that $\trcST{} =_{\vartriangle} {\trcST{}}'$ have
the same center of mass. Note that this center of mass, which is represented by a degenerated
triangle, only has one representation. Second, all corner points representing a \st triangle are
\cotemp. Therefor, we can conclude that the cross-ratio of the time coordinates of three triangles
$\trcST{1}$, $\trcST{2}$ and $\trcST{3}$ is the same as the cross-ratio of the time coordinates of
any triple of triangles ${\trcST{1}}'$, ${\trcST{2}}'$ and ${\trcST{3}}'$, such that $\trcST{l}
=_{\vartriangle} {\trcST{l}}' (1 \leq l \leq 3)$. It now follows from the first and second
statements, that given the \st triangles $\trcST{1}$, $\trcST{2}$, $\trcST{3}$, $\trcST{4}$,
$\trcST{5}$ and $\trcST{6}$, $$\predi{Cas}(\trcST{1}, \ab\trcST{2}, \ab\trcST{3}, \ab\trcST{4},
\ab\trcST{5}, \ab\trcST{6}) \leftrightarrow \predi{Cas}({\trcST{1}}', \ab{\trcST{2}}',
\ab{\trcST{3}}', \ab{\trcST{4}}', \ab{\trcST{5}}', \ab{\trcST{6}}'),$$ for any ${\trcST{l}}'$ such
that $\trcST{l} =_{\vartriangle} {\trcST{l}}' (1 \leq l \leq 6)$.

Now we have to prove that the composed formulas always return consistent triangle relations. Let
$\hat{\varphi}$ and $\hat{\psi}$ be two formulas in \fom{$\{\posCotemp,\ab \beforetrs,
\ab\predi{Cas} \}$}{$\sigmatrST$}, of arity $k_{\varphi}$ and $k_{\psi}$ respectively, already
defining consistent triangle relations. Then, the formula $(\hat{\varphi} \land \hat{\psi})$
(resp., $(\hat{\varphi} \lor \hat{\psi})$) also defines a triangle relation. This follows from the
fact that the free variables of $(\hat{\varphi}\land\hat{\psi})$ (resp., $(\hat{\varphi} \lor
\hat{\psi})$) are free variables in $\hat{\varphi}$ or $\hat{\psi}$. The universe of all triangles
is trivially consistent. If a consistent subset is removed from this universe, the remaining part
is still consistent. Therefor, $\neg\hat{\varphi}$ is well-defined. Finally, because consistency is
defined argument-wise, the projection $\exists\trcST{1}\,\hat{\varphi}(\trcST{1}, \trcST{2},
\ldots, \trcST{k_{\varphi}})$ is consistent. \qed

\medskip

\begin{theorem}[\textbf{Expressiveness of  FO($\{\posCotemp,\ab \beforetrs, \ab\predi{Cas}
\}$)}]\rm\label{theorem-sttr1-soundcomplete} Let $\sigmatrST$ be a database schema.
 The language \fom{$\{\posCotemp,\ab \beforetrs, \ab\predi{Cas} \}$}{$\sigmatrST$} is sound and complete for the $({\cal AC}_{st}, {\cal A}_t)$-generic
 \fok-queries on \st triangle databases.
\end{theorem}

As usual, we prove this theorem using the following two lemma's:

\begin{lemma}[\textbf{Soundness  \foms{$\{\posCotemp,\ab \beforetrs, \ab\predi{Cas}\}$}}]\rm\label{lemma-sttr1-sound}
 Let $\sigmatrST$ be a \\ spa\-tio-\-tem\-po\-ral triangle database
schema. Then \fom{$\{\posCotemp,\ab \beforetrs, \ab \predi{Cas} \}$}{$\sigmatrST$} is sound
for the $({\cal AC}_{st}, {\cal A}_t)$-generic \fok-queries on \st triangle databases.
\end{lemma}

\par\noindent{\bf Proof.} Let $\sigmatrST = \{\RtrST_1, \RtrST_2, \ldots, \RtrST_m\}$ be a \st triangle database schema. Similar to
the proof of Lemma~\ref{lemma-tr-sound}, this proof consists of two parts.

First, let $\sigmaptST = \{\RptST_1, \RptST_2, \ldots, \RptST_m\}$ be a \st point database sche\-ma
where the arity of $\RptST_i$ is $3\times ar(\RtrST_i)$, for $i = 1,2, \ldots, m$. We show that
each formula of \fom{$\{\posCotemp,\ab \beforetrs, \ab\predi{Cas} \}$}{$\sigmatrST$} can be
translated in \fobSTsigma. We to this by induction on \fom{$\{\posCotemp,\ab \beforetrs,
\ab\predi{Cas} \}$}{$\sigmatrST$}-formulas. Next, we have to prove that each \fom{$\{\posCotemp,\ab
\beforetrs, \ab\predi{Cas} \}$}{$\sigmatrST$}-query defines a consistent \st triangle relation.

\medskip

We start with the first part of this proof. Let $\RptST_i (1 \leq i \leq m)$ be the corresponding
\st point relation names of arity $3\times ar(\RtrST_i)$ and let $\sigmaptST$ be the \st (point)
database schema $\{\RptST_1, \RptST_2, \ldots, \RptST_m\}$. Let $\hat{\varphi}$ be a
\fom{$\{\posCotemp,\ab \beforetrs, \ab\predi{Cas}$}{$\sigmatrST \}$}-formula.

Each triangle variable $\trST{}$ in $\hat{\varphi}$ is translated naturally by three \st point
variables $u_1, u_2, u_3$. As we assume that all points composing a  \st triangle are \cotemp, we
add the formula $$\predi{Cotemp}(u_1, u_2) \land \predi{Cotemp}(u_2, u_3)$$ to the beginning of the
translation of the sub-formula where $\trST{}$ appears first.  In the remainder of this proof we
will omit these temporal constraints to keep formulas shorter and hence more readable, but always
assume them.

The formulas in \fom{$\{\posCotemp,\ab \beforetrs, \ab\predi{Cas} \}$}{ $\sigmatrST$} are build
from atomic formulas, composed by the operators $\land$, $\wedge$ and $\neq$ and quantification.
The atomic formulas of \fom{$\{\posCotemp,\ab \beforetrs, \ab\predi{Cas} \}$}{$\sigmatrST$} are
equality constraints between \st triangle variables, the triangle predicates $\posCotemp$,
$\beforetrs$ and $\predi{Cas}$ applied to \st triangle variables, and predicates of the form
$\RtrST_i(\trST{1},\ab \trST{2}, \ldots, \ab\trST{ar(\RtrST_i)})\ab (1 \leq i \leq m)$, where
$\RtrST_i \in \sigmatrST$. As this proof is analogous to the proof of Lemma~\ref{lemma-tr-sound},
we only give the translation of the atomic formulas:

\begin{enumerate}[(i)]
    \item The translation of $(\trST{1} = \trST{2})$ is
    $$\displaylines{\qquad \bigvee_{\sigma(1,2,3) = (i_1, i_2, i_3), \sigma \in {\cal S}_3} (u_{1,1} \mathrel{=} u_{2,{i_1}}\land u_{1,2} \mathrel{=} u_{2,{i_2}}\land u_{1,3} \mathrel{=} u_{2,{i_3}}),\qquad}$$
where ${\cal S}_3$ is the set of all permutations of $\{1,2,3\}$.
    \item In the proof of Lemma~\ref{lemma-tr-sound}, we already showed
that the predicate $\pos$ can be expressed in \foms{$\{\betws\}$}.
    \item Expressions of the form $\beforetr{\trST{1}}{\trST{2}}$ are translated as follows:
    $$\displaylines{\qquad\before{u_{1,1}}{u_{2,1}}.\qquad}$$ Recall that the formulas expressing
    that the corner points of each triangle should be \cotemp are already added to the translation.

    \item For the predicate $\predi{Cas}$, first we need to to express in
    \fom{$\{\betwsCotemp,\ab\befores, \ab\eqcrsts \}$}{$\sigmaptST$}
    that some point (in  $\RnR{2}$) is the center of mass of a triangle, represented by three other points, all \cotemp with the first point.

    \begin{figure}
  \centerline{\includegraphics[width=175pt]{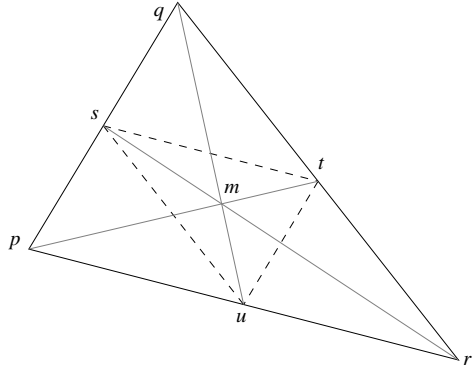}}
  \caption{The center of mass of the triangle $pqr$ is the intersection of the medians $pt$, $qu$ and $rs$. Also, the lines $tu$, $us$ and $st$ are parallel to $pq$, $qr$ and $rp$, respectively.}\label{fig-centerom}
\end{figure}

Figure~\ref{fig-centerom} illustrates the construction of the center of mass of a triangle. Given a
triangle $\trcST{pqr}$. There is only one way of constructing a triangle $\trcST{stu}$ inscribed in
$\trcST{pqr}$ such that each side of $\trcST{stu}$ is parallel to a side of $\trcST{pqr}$. The
corner points of $\trcST{stu}$ are in the middle of the sides of $\trcST{pqr}$. Hence, the center
of mass of $\trcST{pqr}$ is the intersection of the line segments connecting the corner points of
$stu$ with the opposite corner point of $\trcST{pqr}$. The next formula expresses the predicate
$\predi{CenterOM}$ in the language \fobST. The free variables are $v$ (representing the center of
mass), $u_1$, $u_2$ and $u_3$ (representing the corner points of the triangle).

$$\displaylines{\qquad \exists w_1\,\exists w_2\,\exists w_3\,(\betwCotemp{u_1}{w_1}{u_2} \land \betwCotemp{u_2}{w_2}{u_3} \land \splits\betwCotemp{u_3}{w_3}{u_1} \land \predi{Par}(u_1, u_2, w_2, w_3) \land\predi{Par}(u_2, u_3, w_1, w_3) \land \splits\predi{Par}(u_3,u_1,w_1, w_2)\land \betwCotemp{u_1}{v}{w_2} \land \splits\betwCotemp{u_2}{v}{w_3} \land \betwCotemp{u_3}{v}{w_1}).\qquad}$$

Here, $\predi{Par}(v_1, v_2, v_3, v_4)$ is an abbreviation for the sub-formula
$$\displaylines{\qquad\neg\exists w\,(\predi{Collinear}(w, v_1, v_2) \land \predi{Collinear}(w,
v_3, v_4)).\qquad}$$ We now give the expression translating $\predi{Cas}(\trST{1}, \ab\trST{2},
\ab\trST{3}, \ab\trST{4}, \trST{5}, \trST{6})$. The following formula has $(6\times 3)$ free point
variables $u_{1,1} u_{1,2}, u_{1,3}, \ab u_{2,1}, u_{2,2}, u_{2,3}, \ab\ldots, \ab u_{6,1},
u_{6,2}, u_{6,3}$ that are the translation of the triangle variables $\trST{1}, \ab\trST{2},
\ab\ldots, \ab\trST{6}$.
$$\displaylines{\qquad\exists v_1\,\exists v_2\,\exists v_3\,(\splits\bigwedge_{i=1}^3 \predi{CenterOM}(v_i, u_{i,1}, u_{i,2}, u_{i,3}) \land \eqcrst{v_1}{v_2}{v_3}{u_{4,1}}{u_{5,1}}{u_{6,1}}).\qquad}$$
\item The translation of a formula of the type $\RtrST(\trST{1}, \trST{2}, \ldots, \trST{k})$,
where $\RtrST \in \sigmatrST$ is $$\Rpt(u_{1,1}, u_{1,2}, u_{1,3}, u_{2,1}, u_{2,2},
u_{2,3},\ldots, u_{k,1}, u_{k,2}, u_{k,3}).$$ The correctness of this translation follows from
Definition~\ref{def-triangle-base} and Remark~\ref{remark-triangle-can}. \qed
\end{enumerate}

\medskip

We can also show the possibility of the translation in the other direction. As the proof of
Lemma~\ref{lemma-sttr1-complete} is completely analogous to the proof of
Lemma~\ref{lemma-tr-complete}, we omit it. The only new items are the translations of the \st point
predicates $\befores$ and $\eqcrsts$ into \foms{$\{\posCotemp,\ab \beforetrs, \ab\predi{Cas} \}$}.
It is easy to see that these translations involve only replacing point variables by triangle
variables that represent points.

\begin{lemma}[\textbf{Completeness of FO($\{\posCotemp,\ab \beforetrs, \ab\predi{Cas}
\}$)}]\rm\label{lemma-sttr1-complete} Let $\sigmatrST$ be a \st triangle database schema. The
language \fom{$\{\posCotemp,\ab \beforetrs, \ab\predi{Cas} \}$}{$\sigmatrST$} is complete for the
$({\cal AC}_{st}, {\cal A}_t)$-generic \fok-queries on \st triangle databases.\qed
\end{lemma}

We now propose an alternative language, with the same expressiveness as the language
\foms{$\{\posCotemp,\ab \beforetrs, \ab\predi{Cas} \}$}, which allows us to talk about areas of
triangles.

\subsubsection{Expressiveness of the Language \foms{$\{\posCotemp,\ab \beforetrs, \ab\predi{Lex}
\}$}}

We start this subsection with some geometric constructions. We will use those to express the
predicate $\predi{Lex}$ in the language \foms{$\{\betwsCotemp,\ab\befores, \ab\eqcrsts \}$}. For
these constructions, we assume that all \st points and triangles are \cotemp.

\begin{figure}
  \centerline{\includegraphics[width=150pt]{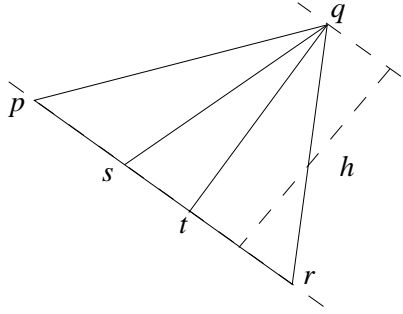}}
  \caption{The area of $\trc{pqr}$ is to the area of $\trc{pqs}$ as the length of $pr$ to the length of $ps$.}\label{fig-lex2}
\end{figure}

\begin{observation}\rm\label{obs-lex-1}
Let two triangles $\trcST{pqr}$ and $\trcST{pqs}$ be given. If the point $s$ is chosen on the line
segment $pr$ such that the cross ration of $p$, $s$ and $r$ equals $c$, then the areas of
$\trcST{pqr}$ and $\trcST{pqs}$ have a ratio which is also equal to $c$. The correctness of this
construction is easy to verify because the area of a triangle is half the length of its base line
multiplied by its height. As $\trcST{pqr}$ and $\trcST{pqs}$ have both height $h$, their areas have
the same relation as the lengths of their base lines $ps$ and $pq$. Figure~\ref{fig-lex2}
illustrates this observation.

Suppose we have three triangles $\trcST{pqr}$, $\trcST{pqs}$ and $\trcST{pqt}$, such that the
points $q$, $r$, $s$ and $t$ are all collinear (suppose they are arranged as in
Figure~\ref{fig-lex2}). Then it is true that $$\frac{A(\trcST{pqt}) - A(\trcST{pqs})}{
A(\trcST{pqr})- A(\trcST{pqs})} = \frac{\mid st \mid}{\mid sr \mid}.$$  \end{observation}

\medskip

So it turns out to be possible to convert area ratios to cross-ratios of collinear points, for
triangles that have the special configuration as described in Observation~\ref{obs-lex-1}. We will
observe next that it is possible, given three triangles $\trcST{1}$, $\trcST{2}$ and $\trcST{3}$
such that $\trcST{1}$ is part of $\trcST{2}$ and $\trcST{2}$ part of $\trcST{3}$, to construct
triangles $\trcST{4}$ and $\trcST{5}$ with the same area as $\trcST{1}$ and $\trcST{2}$,
respectively, such that $\trcST{4}$, $\trcST{5}$ and $\trcST{3}$ have this special configuration.

\begin{figure}
  \centerline{\includegraphics[width=275pt]{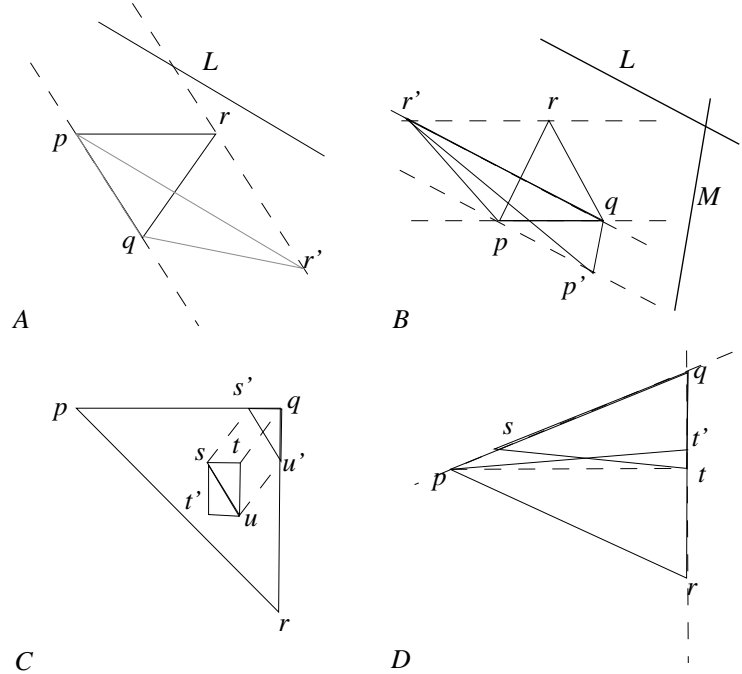}}
  \caption{Area-preserving affine-invariant constructions.}\label{fig-lex}
\end{figure}

\begin{observation}\rm\label{obs-lex-2}
Given a pair of triangles $\trcST{1}$ and $\trcST{2}$ such that $\trcST{1}$ is part of $\trcST{2}$.
Following the construction steps described below, we can construct a triangle $\trcST{3}$, with the
same area as $\trcST{1}$. The triangle $\trcST{3}$ shares one side with $\trcST{2}$ and its third
corner point is on one of the other sides of $\trcST{2}$.

\medskip
\par\noindent{\it Construction step 1:}
\medskip
\par\noindent Given a triangle $\trcST{pqr}$ and a line $L$, we construct a triangle with the same
area as $\trcST{pqr}$, but one side parallel to $L$. We do this by  moving the point $r$ over the
line through $r$ and parallel with $pq$ until one of the line segments $pr$ or $qr$ is parallel to
$L$. The resulting triangle $\trcST{pqr'}$ has the same area as $\trcST{pqr}$ because it has the
same base line segment and the same height as $\trcST{pqr}$. Figure~\ref{fig-lex}, part $A$,
illustrates this construction.

If we apply this construction twice, we can construct a triangle with two sides parallel to two
given (different) lines. This is shown in Figure~\ref{fig-lex}, part $B$, where the triangle
$\trcST{pqr}$ is first transformed into $\trcST{pqr'}$ and, in a second step, into $\trcST{p'qr'}$.

\medskip
\par\noindent{\it Construction Step 2:}
\medskip
\par\noindent Let a triangle $\trcST{pqr}$ and a smaller triangle, either $\trcST{stu}$ or $\trcST{st'u}$, which has two sides
parallel to the sides $pq$ and $qr$ respectively of $\trcST{pqr}$, be given. There are two possible
orientations for the smaller triangle. Either it is oriented in such a way that the corner point
$t'$ is on the opposite side of $su$ than the point $q$, as is the case for triangle $\trcST{st'u}$
in Figure~\ref{fig-lex}, part C, or it is oriented otherwise, as is the case for triangle
$\trcST{stu}$. In the first case, we \emph{flip} $\trcST{st'u}$ by constructing the parallelogram
$st'ut$, and then considering the triangle $\trcST{stu}$.

Next, starting from a triangle $\trcST{stu}$ with the right orientation, we construct a triangle
$\trcST{qs'u'}$ which has the same area as $\trcST{stu}$, but shares a corner point with
$\trcST{pqr}$ and has its other corner points on the two sides of $\trcST{pqr}$, adjacent to the
common corner point. This transformation involves only a translation, which can be carried out by
constructing a set of parallel lines.

\medskip
\par\noindent{\it Construction Step 3:}
\medskip
\par\noindent Given a triangle $\trcST{pqr}$, and a triangle $\trcST{sqt}$ such that $s$ lies on the line
through $pq$ and $t$ lies on the line through $qr$. We can construct a triangle $\trcST{pqt'}$ that
has the same area as $\trcST{sqt}$ by making sure that the cross-ratio of the points $p$, $s$ and
$q$ equals the cross-ratio of the points $t$, $t'$ and $q$. Figure~\ref{fig-lex}, part $D$,
illustrates this construction.

Using the above three steps, we constructed, starting from two arbitrary triangles, one being part
of the other, two triangles that have the desired configuration. \end{observation}

We now can prove that our alternative language, \foms{$\{\posCotemp, \ab\befores, \ab\predi{Lex}
\}$} also is sound and complete for the $({\cal AC}_{st}, {\cal A}_t)$-generic \fok-queries on
triangle databases. As the proof is completely analog as the proof of
Theorem~\ref{theorem-sttr1-soundcomplete}, except for the translations of the predicates
$\predi{Lex}$ and $\eqcrsts$, we only give those translations.

\begin{theorem}[\textbf{Expressiveness of FO($\{\posCotemp,\ab \beforetrs, \ab\predi{Lex}
\}$)}]\rm\label{theorem-sstr2-soundcomplete} Let $\sigmatrST$ be a \st triangle database schema.
 The language \fom{$\{\posCotemp,\ab \beforetrs, \ab\predi{Lex} \}$}{ $\sigmatrST$} is sound and complete for the $({\cal AC}_{st}, {\cal A}_t)$-generic
 \foms{\sigPolyNoBr}-queries on \st triangle databases.
\end{theorem}
\par\noindent{\bf Proof.}
First, let $\sigmatrST = \{\RtrST_1, \RtrST_2, \ldots, \RtrST_m\}$ be a \st triangle database
schema and let $\sigmaptST = \{\RptST_1, \RptST_2, \ldots, \RptST_m\}$ be a \st point database
schema where the arity of $\RptST_i$ is $3\times ar(\RtrST_i)$, for $i = 1,2, \ldots, m$.

We first show that the predicate $\predi{Lex}$ can be expressed in \fobSTsigma. We verify that this
predicate is invariant for transformations in $({\cal AC}_{st}, {\cal A}_t)$. The proportion of the
areas of two co-temporal triangles is invariant under affinities. This, together with the fact that
cross-ratios of time moments are invariant under affine transformations of the time, shows that the
predicate $\predi{Lex}$ is $({\cal AC}_{st}, {\cal A}_t)$-invariant.

\begin{figure}
  \centerline{\includegraphics[width=150pt]{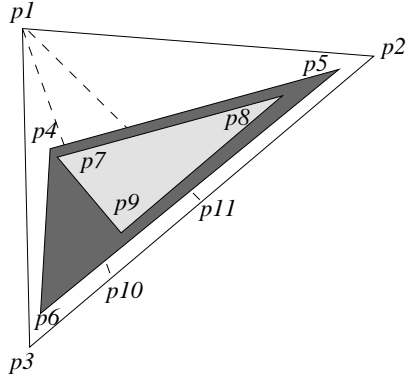}}
  \caption{An illustration of the predicate $\predi{SameRelArea}$. The expression $\predi{SameRelArea}(p_1, p_2, \ldots, p_{11})$
  will be true if and only if two conditions are met. First, the triangle with corner points $p_{7}$, $p_{8}$ and $p_{9}$
  (the light shaded one) is part of the triangle with corner points $p_{4}p$, $p_{5}$ and $p_{6}$ (the dark shaded one),
  which is part of the triangle with corner points $p_1$, $p_2$ and $p_3$ (the white triangle).
  Second, The areas of the light and dark shaded triangles are to the area of the white triangle as the areas of the
  triangles with corner points $p_1$, $p_2$ and $p_{10}$, resp. $p_1$, $p_2$ and $p_{11}$ to the area of the white triangle.}\label{fig-samearea}
\end{figure}

The constructions described in Observation~\ref{obs-lex-2} can all be expressed in the language
\fobST. They mainly involve parallelism-constraints on points.

Let $\predi{SameRelArea}$ be the abbreviation for a predicate in \fobST of arity $11$. The first
nine free variables represent the corner points of three \cotemp triangles, such that the first
triangle is part of the second, which is again part of the third triangle. The two last point
variables are located on one side of the third triangle, in such a way that the parts they define
of the third triangle (denoted triangle four and five), are part of each other also. Finally, the
proportion of the areas of the first three triangles is the same as the proportion of the areas of
the fourth, fifth and third triangle. Fig~\ref{fig-samearea} illustrates this predicate.

The translation of $\predi{Lex}(\trST{1}, \trST{2}, \ldots, \trST{6})$ then is the following
expression:

$$\displaylines{\qquad \exists v_1\,\exists v_2\,(\predi{SameArea}(u_{1,1}, u_{1,2}, u_{1,3}, u_{2,1}, u_{2,2}, u_{2,3}, u_{3,1}, u_{3,2}, u_{3,3}, v_1, v_2) \land \splits \eqcrst{v_1}{v_2}{u_{3,3}}{u_{4,1}}{u_{5,1}}{u_{5,2}} ) ,\qquad}$$
if $\trST{i}$ is translated by $u_{i,1}$, $u_{i,2}$ and $u_{i,2}$ for $i = 1, \ldots, 6$.

The translation in the other direction is simpler. The formula
$\eqcrst{u_1}{u_2}{u_3}{u_4}{u_5}{u_6}$ can be expressed as
$$\displaylines{\qquad \exists\trST{7}\,\exists\trST{8}\,\exists\trST{9}\,\exists\trST{10}\,\exists\trST{11}\,(\splits\predi{CornerP}(\trST{7}, \trST{8}, \trST{1}, \trST{9}) \land \predi{CornerP}(\trST{7}, \trST{8}, \trST{2}, \trST{10}) \splits
\land \predi{CornerP}(\trST{7}, \trST{8}, \trST{3}, \trST{11}) \land \predi{Lex}(\trST{9},
\trST{10}, \trST{11}, \trST{4}, \trST{5}, \trST{6})).\qquad}$$ \qed

\subsection{Physics-based Classes}\label{sec-physics-predi}

In the previous section, we investigated a triangle language for $({\cal AC}_{st}, {\cal
A}_t)$-generic triangle queries.  Next, we focus on triangle languages for the physics-based
queries, \ie, those generic for the group $({\cal V}_{st}, {\cal A}_t)$ of velocity-preserving
transformations and the group $({\cal AC}_{st}, {\cal A}_t)$ of acceleration-preserving
transformations.

In \cite{ghk-01}, the query languages expressing queries generic for the physics-based
transformation groups were found by starting with the languages expressing the affine-invariant
spatial point queries. The reason was that the physics-based transformation groups of $\RnR{2}$ are
a subgroup of the affinities of $\Rn{3}$, and that \st points in $\RnR{2}$ can be interpreted
equally well as points in $\Rn{3}$.

Here, it is not expedient to do so. We can see \st triangles in $\RnR{2}$ as convex objects in
$\Rn{3}$, but then the predicate $\pos$ would not make much sense, as \st triangles can only
overlap when they exist at the same moment in time. Another solution would be to choose other
convex objects, that have a temporal extend of more than one time moment. But, these objects would
make rather poor \st objects. Indeed, even if all corner points of a triangle in $\Rn{2}$ move with
a linear function of time, this movement can result in a \ndim{3} object bounded by non-planar
surfaces, and hence possibly not convex.

Therefor, we take another approach and start with the predicates $\posCotemp$ and $\beforetrs$, as
in the previous section, and add other predicates until the resulting language is expressive
enough. In concrete, this means that we have to be able to translate the point predicate $\betws$
in that language.

As $({\cal V}_{st}, {\cal A}_t) \subset ({\cal AC}_{st}, {\cal A}_t)$, we start with the
acceleration preserving transformations first, and later extend the language expressing all $({\cal
AC}_{st}, {\cal A}_t)$-generic queries in such a way we obtain a language expressing the $({\cal
V}_{st}, {\cal A}_t)$-generic queries.

\subsubsection{$({\cal AC}_{st}, {\cal A}_t)$-generic Queries}

For the acceleration-preserving queries, we introduce the \st triangle predicate $\predi{SAS}$
(which is an abbreviation for ``Same Average Speed''). Let $\trST{1}, \trST{2}, \trST{3}$ and
$\trST{4}$ be four triangles that have center of mass $p_i = (a_i, b_i, \tau_i), i = 1 \ldots 4$.
Furthermore, $\tau_1 \leq \tau_2$ and $\tau_3 \leq \tau_4$. Then $\predi{SAS}(\trST{1}, \trST{2},
\trST{3}, \trST{4})$ is true if and only if $$\frac{a_2 - a_1}{\tau_2 - \tau_1} = \frac{a_4 -
a_3}{\tau_4 - \tau_3} \textrm{ and } \frac{b_2 - b_1}{\tau_2 - \tau_1} = \frac{b_4- b_3}{\tau_4 -
\tau_3}.$$ In other words, the movement from $\trST{1}$ to $\trST{2}$ has the same average speed,
in both $x$- and $y$-direction, as the movement from $\trST{3}$ to $\trST{4}$.

We now show that the language \foms{$\{\posCotemp, \beforetrs, \predi{SAS}\}$} is sound and
complete for the $({\cal AC}_{st}, {\cal A}_t)$-generic \foms{\sigPolyNoBr}-queries on triangle
databases.

As soundness and completeness proof are completely analogous to those of the previous section, we
only give the translations of the triangle predicates from the set $\{\posCotemp, \ab \beforetrs,\ab
\predi{SAS}\}$ into \foms{$\{\betws\}$} and of the point predicate $\betws$ into the logic 
\foms{$\{\posCotemp, \beforetrs, \predi{SAS}\}$}.

\begin{theorem}[\textbf{Expressiveness of  FO($\{\posCotemp,\ab \beforetrs, \ab\predi{SAS}
\}$)}]\rm\label{theorem-acc-soundcomplete} Let $\sigmatrST$ ba a triangle database schema. Let
$\overline{\sigma}^{st}$ be the corresponding semi-algebraic database schema. The language
\fom{$\{\posCotemp,\ab \beforetrs, \ab\predi{SAS} \}$}{$\sigmatrST$} is sound and complete for the
$({\cal AC}_{st}, {\cal A}_t)$-generic \fom{\sigPolyNoBr}{$\overline{\sigma}^{st}$}-queries on
triangle databases over $\sigmatrST$.
\end{theorem}
\par\noindent{\bf Proof sketch.}
  Let $\sigmatrST = \{\RtrST_1, \RtrST_2, \ldots, \RtrST_m\}$ be a
spatial triangle database
    schema.
Let $\RptST_i, 1 \leq i \leq m$ be the corresponding spatial point relation names of
    arity $3\times ar(\RtrST_i)$ and let $\sigmaptST$ be the spatial database schema $\{\RptST_1, \RptST_2, \ldots, \RptST_m\}$.
    Let $\overline{R}^{st}_i, 1 \leq i \leq m$ be the corresponding constraint relation names of
    arity $6\times ar(\RtrST_i)$ and let $\overline{\sigma}^{st}$ be the spatial database schema $\{\overline{R}^{st}_1, \overline{R}^{st}_2, \ldots, \overline{R}^{st}_m\}$.

In this proof sketch, we only give the translation of $\predi{SAS}$ into
\fom{$\{\betws\}$}{$\sigmaptST$}. For the translations of $\posCotemp$ and $\beforetrs$, see
Section~\ref{sec-tr-logic-spatial} and Section~\ref{sec-st-predi} respectively.

Given the expression $\predi{SAS}(\trST{1}, \trST{2}, \trST{3}, \trST{4})$. The following formula
is its translation into \fom{$\{\betws\}$}{$\sigmaptST$}:

$$\displaylines{\qquad \exists v_1\,\exists v_2\,\exists v_3\,\exists v_4\,(\splits\bigwedge_{i=1}^4 \predi{CenterOM}(v_i, u_{i,1}, u_{i,2}, u_{i,3}) \land \before{v_1}{v_2} \land \before{v_3}{v_4}\splits\land \predi{CoPlanar}(v_1, v_2, v_3, v_4)\land\neg\exists w\,(\predi{Collinear}(w, v_1, v_2) \land \predi{Collinear}(w, v_3, v_4)) ).\qquad}$$

We have omitted the sub formulas expressing that the corner points of a triangle should be \cotemp.
The predicate $\predi{CoPlanar}$ expresses that four \ndim{3} points are co-planar. It is clear
that this is an affine invariant and \fok-expressible.

For the definition of $\predi{CenterOM}$, we refer to the proof of Lemma~\ref{lemma-sttr1-sound}.

We next prove that the predicate $\betws$ can be expressed in  \fom{$\{\posCotemp,\ab
\beforetrs, \ab\predi{SAS} \}$}{$\sigmatrST$}. This translation is not complicated. If the
expression $\betw{p}{q}{r}$ holds for three points $p$, $q$ and $r$, then either they are all
\cotemp or they all exist at a different time moment. In the first case, we can translate $\betws$
using $\pos$, as we showed in the proof of Lemma~\ref{lemma-tr-complete}. If they all have a
different time coordinate, we can express that $q$ is between $p$ and $r$ using $\predi{SAS}$:

$$\displaylines{\qquad(\predi{CoTemp}(\trST{1}, \trST{2}) \land \predi{CoTemp}(\trST{2}, \trST{3}) \land \splits \betwtr{\tr{p}}{\tr{q}}{\tr{r}}) \lor ( \predi{SAS}(\trST{1}, \trST{2}, \trST{2}, \trST{3})).\qquad}$$

 In the previous formula, we have omitted the sub-formulas expressing that the triangles translating the point
variables should be points. \qed

\medskip

Since the group $({\cal V}_{st}, {\cal A}_t)$ is a subgroup of the group $({\cal AC}_{st}, {\cal
A}_t)$, we use our knowledge from this subsection to extend the language \fom{$\{\posCotemp,\ab
\beforetrs, \ab\predi{SAS} \}$}{$ \sigmatr$}, which we will do next.

\subsubsection{$({\cal V}_{st}, {\cal A}_t)$-generic Queries}

In this subsection, we propose a language sound and complete of the first-order $({\cal V}_{st},
{\cal A}_t)$-generic triangle queries. We add the element $\predi{NoSp}$ (an abbreviation for ``No
Speed'') to the set $\{\posCotemp, \ab \beforetrs, \ab \predi{SAS}\}$.

Suppose two \st triangles $\trcST{1}$ and $\trcST{2}$  have center of mass $p_i = (a_i, \ab b_i,
\ab\tau_i),\ab i = 1, 2$. If we furthermore assume that $\tau_1 \leq \tau_2$, then
$\predi{NoSp}(\trST{1}, \trST{2})$ is true if and only if $a_1 = a_2$ and $b_1 = b_2$. In other
words, the average speed is zero, both triangles are on the same position.

We now show that the language \foms{$\{\posCotemp, \beforetrs, \predi{SAS}, \predi{NoSp}\}$} is
sound and complete for the $({\cal AC}_{st}, {\cal A}_t)$-generic \foms{\sigPolyNoBr}-queries on
triangle databases.

As soundness and completeness proof are completely analogous to those of the previous section, we
only give the new translations.

\begin{theorem}[\textbf{Expressiveness of FO($\{\posCotemp,\ab \beforetrs,
\ab\predi{SAS}, \predi{NoSp}\}$)}]\rm\label{theorem-vel-soundcomplete} Let $\sigmatrST$ be a
triangle database schema. Let $\overline{\sigma}^{st}$ be the corresponding semi-algebraic database
schema. Then the language \fom{$\{\posCotemp,\ab \beforetrs, \ab\predi{SAS}, \predi{NoSp} \}$}{$
\sigmatrST$} is sound and complete for the $({\cal AC}_{st}, {\cal A}_t)$-generic
\fom{\sigPolyNoBr}{$\overline{\sigma}^{st}$}-queries on triangle databases over $\sigmatrST$.
\end{theorem}
\par\noindent{\bf Proof sketch.}
 Let $\sigmatrST = \{\RtrST_1, \RtrST_2, \ldots, \RtrST_m\}$ be a
spatial triangle database
    schema. Let $\RptST_i, \ab 1 \leq i \leq m$ be the corresponding spatial point relation names of
    arity $3\times ar(\RtrST_i)$ and let $\sigmaptST$ be the spatial database schema $\{\RptST_1, \RptST_2, \ldots, \RptST_m\}$.
    Let $\overline{R}^{st}_i, \ab 1 \leq i \leq m$ be the corresponding constraint relation names of
    arity $6\times ar(\RtrST_i)$ and let $\overline{\sigma}^{st}$ be the spatial database schema $\{\overline{R}^{st}_1, \overline{R}^{st}_2, \ldots, \overline{R}^{st}_m\}$.

In this proof sketch, we only give the translation of the predicate $\predi{NoSp}$ into the language 
\foms{$\{\betws,\befores, \predi{EqSpace} \}$} and of the predicate $\predi{EqSpace}$ into the
language \foms{$\{\posCotemp, \beforetrs, \predi{SAS}, \predi{NoSp}\}$}.

The next formula, with free variables $u_1, u_2, u_3, v_1, v_2, v_3$ is the translation of
$\predi{NoSp}(\tr{u}, \tr{v})$ into \fom{$\{\betws,\befores, \predi{EqSpace} \}$}{$\sigmatr$}.

\medskip

\noindent$\exists w_1\exists w_2(\predi{CenterOM}(w_1, u_1,u_2,u_3) \land \predi{CenterOM}(w_2,
v_1,v_2,v_3) \land \predi{EqSpace}(w_1,w_2)).$

\medskip

Finally, the formula

$$\displaylines{\qquad\predi{Point}(\trST{u}) \land\predi{Point}(\trST{v}) \land \predi{NoSp}(\trST{u},
\trST{v})\qquad}$$

translates $\predi{EqSpace}(u,v)$ into \foms{$\{\posCotemp, \beforetrs, \predi{SAS},
\predi{NoSp}\}$}. Note that, if a triangle is degenerated into a point, its center of mass is equal
to the triangle itself. \qed

We end with a note on safety of \st triangle database queries.

\subsection{Safety of Spatio-temporal Triangle Database Queries}\label{sec-safety-st}

In Section~\ref{sec-safety}, we addressed the safety-problem for spatial triangle queries. In the
spatial case, we defined a query to be safe when it returns a finite number of triangles on an
input consisting of a finite number of triangles. Due to our choice of not considering convex
objects in \ndim{(2+1)} space  but \st triangles as basic objects for our language (see
Remark~\ref{remark-generalization} and the start of Section~\ref{sec-physics-predi}), this
definition does not carry over to the \st case. Indeed, it would be very unnatural to consider \st
databases containing a finite number of \st triangles only.

It follows from a well-known property of \sa sets that there exists a finite partition of the time
domain of a \st database in points and open intervals such that within such an interval all
snapshots are isotopic to each other and there exists a continuous family of homeomorphisms mapping
these snapshots to each other (this is explained in more detail in~\cite{kpv-00}). So, \st
databases that are \sa sets can in fact be considered ``finite'' \st databases in general. However,
given a \st relation $R$, a formula in \foms{\sigPolyNoBr, $R$} that expresses this partition for
$R$ does not exist. The partition can be computed by performing a CAD (Cylindrical Algebraic
Decomposition)~\cite{collins}.

A desirable property for a ``finite'' \st triangle database, would be that every snapshot of the
\st database can be represented using a finite number of \st triangles. This essentially is the
requirement that each snapshot would be a finite spatial triangle relation. It is easy to see that
we can express this requirement using $\posCotemp$, using the results of Section~\ref{sec-safety}.

We can conclude that the safety problem for \st triangle databases is strongly related to the
safety problem for spatial triangle databases. Because we do not consider real \st objects as basic
objects for our language and as basic elements of \st triangle databases, we can only ask that each
snapshot of a \st triangle database is finite.

\section{Conclusion}
In this article, we introduced the new triangle-based query language \fod. The use of
triangles instead of points or real numbers is motivated by the spatial (spatio-temporal) practice,
where data is often represented as a collection of (moving) triangles.

We showed that our query language has the same expressiveness as the affine-invariant  \fob\/-queries on triangle databases. We did this by showing that our language is sound and
complete for the \fob\/-queries on triangle databases.

Afterwards, we gave several examples to illustrate the expressiveness of the triangle-based
language and the ease of use of manipulating triangles.

We then turned to the notion of safety. We showed that, although we cannot decide whether a
particular Tquery returns a finite output given a finite input, we can decide whether the output is
finite. We also extended this finiteness to the more intuitive notion of sets that have a finite
representation. We proved that we can decide whether the output of a query has a finite
representation and compute such a finite representation in \fod.

Besides the intuitive manipulation of spatial data represented as a collection of triangles,
another motivation for this language is that it can serve as a first step towards a natural query
language for spatio-temporal data that are collections of {\em moving triangles}.

Geerts, Haesevoets and Kuijpers~\cite{ghk-01} already proposed point-based languages for several
classes of spatio-temporal queries. The data model used there represented a moving two-dimensional
object as a collection of points in three-dimensional space. There exist however, data models that
represent spatio-temporal data as a collection of moving objects (see for
example~\cite{cz-00,amai03}), which is more natural. Hence, a {\em moving triangle}-based language
with the same expressiveness as the spatio-temporal point languages mentioned above would be much
more useful in practice.

\section*{Acknowledgements}
The authors would like to thank Jan Van den Bussche and Floris Geerts for discussions that have 
given rise to a more concise description of the proposed query languages. 

\bibliographystyle{plain}

\end{document}